\documentclass[aps,10pt,prd,letterpaper,preprintnumbers,amsmath,amssymb,floatfix,nofootinbib]{revtex4}

\usepackage{amsmath,amssymb,graphicx,bm}
\usepackage{enumerate}
\usepackage{subfig}
\usepackage{hyperref}
\usepackage{color}

\usepackage{slashed}            

\usepackage{bbm}

\usepackage{bm}

\usepackage{ulem}

\newcommand{\beq}{\begin {equation}}  
\newcommand{\eeq}{\end   {equation}} 
\newcommand{\bea}{\begin {eqnarray}} 
\newcommand{\eea}{\end   {eqnarray}}  
\newcommand{\baa}{\begin {array}   } 
\newcommand{\eaa}{\end   {array}   }     
\newcommand{\bit}{\begin {itemize} }
\newcommand{\eit}{\end   {itemize} }
\newcommand{\be }{\begin {equation}} 
\newcommand{\ee }{\end   {equation}}

\newcommand\matThree[3]{\ensuremath{\begin{pmatrix} #1  \\ #2 \\ #3\end{pmatrix}}}

\begin{document}


\preprint{ACFI-T16-04}

\title{Indirect Detection Imprint of a CP Violating Dark Sector}

\author{Wei Chao$^{1}$\footnote{Email: chao@physics.umass.edu}}
\author{Michael J. Ramsey-Musolf$^{1,2}$\footnote{Email: mjrm@physics.umass.edu}}
\author {Jiang-Hao Yu$^{1}$\footnote{Email: jhyu@physics.umass.edu}}
\affiliation{ $^1 $Amherst Center for Fundamental Interactions, Physics Department, University of Massachusetts Amherst, Amherst, MA 01003, USA,} 
\affiliation{ $^2 $Kellogg Radiation Laboratory, California Institute of Technology, Pasadena, CA 91125 USA.}



\begin{abstract}
We introduce a simple scenario involving fermionic dark matter ($\chi$) and singlet scalar mediators that may account for the Galactic Center GeV $\gamma$-ray excess while satisfying present direct detection constraints. CP-violation in the scalar potential leads to mixing between the Standard Model Higgs boson and the scalar singlet, resulting in three scalars $h_{1,2,3}$ of indefinite CP-transformation properties. This mixing enables s-wave $\chi{\bar\chi}$ annihilation into di-scalar states, followed by decays into four fermion final states. The observed $\gamma$-ray spectrum can be fitted while respecting present direct detection bounds and Higgs boson properties for $m_{\chi} = 60 \sim 80 $ GeV, and $m_{h_3} \sim m_{\chi}$. Searches for the Higgs exotic decay channel $h_1 \to h_3 h_3$ at the 14 TeV LHC should be able to further  probe the parameter region favored by the $\gamma$-ray excess.

\end{abstract}

\maketitle


\section{Introduction}
\label{sec:intro}

Although the presence of dark matter (DM) has been firmly established by numerous observational data {\it via} its gravitational effects, the particle nature of DM remains a mystery.
It is imperative to search for DM in every feasible way: direct detection, indirect detection, collider searches, {\it etc}.
Both direct detection and collider searches have observed null results, which put constraints on particle DM properties. 
On the other hand, indirect detection offers some hints of the particle nature of the DM. 
The annihilation or decays of particle DM  in the galaxies are expected to produce observable fluxes of cosmic rays, such as anti-protons, positrons, gamma rays, and neutrinos.
Of particular interest are gamma rays from the galactic center and the dwarf spheroidal galaxies, which are able to be detected by the Fermi Large Area Telescope (${\it Fermi}$-LAT)~\cite{FermiLAT:2011ab}.

Recent analyses of the ${\it Fermi}$-LAT data by several groups have identified 
an excess of gamma rays with several GeV energy and a nearly spherically-symmetric distribution in the center of the Milky Way, known as the Galactic Center Excess (GCE)~\cite{Goodenough:2009gk,Abazajian:2012pn,Daylan:2014rsa,Calore:2014xka,TheFermi-LAT:2015kwa}. 
Although astrophysical explanations such as millisecond pulsars have been proposed, the DM annihilation explanation of the GCE has generated much recent attention and is being widely explored. 
The reason is that the morphology of the GCE is consistent with   what is   expected from DM annihilation while complying with the observed thermal relic density.

A common and simple class of DM scenarios that can explain the GCE 
is the two-body DM annihilation $\chi\chi \to f\bar f$, where 
$f$ represents a Standard Model (SM) fermion.
The spectrum of the GCE has been fit well by the DM annihilation  into $\bar b b$
with $m_\chi$ around $31 \sim 40$ GeV and the thermal averaged cross section $\langle \sigma v \rangle_{b\bar b} \sim {\mathcal O}(1\sim 3) \times 10^{-26} {\rm cm}^3/{\rm s}$~\cite{Daylan:2014rsa, Berlin:2014tja}. 
Dark matter annihilation into $\bar \tau \tau$ provides a acceptable fit to the spectrum with lighter $m_\chi$ around 10 GeV and smaller annihilation cross section.
In these models, the DM $\chi$ interacts with the SM fermion through a mediator $\phi$, which could be either scalar, fermion, or vector boson. 
There are two types of scenarios  that explain the GCE through $2\to 2$ annihilation : $s$-channel models with a neutral mediator $\phi$~\cite{Berlin:2014tja}, and $t$-channel models with a charged mediator $\phi$~\cite{Agrawal:2014una, Yu:2014pra}.
Usually in both models a heavy mediator $\phi$ is needed to avoid the LHC constraints, and there should be a mechanism to suppress the spin-independent  DM-nucleus scattering   cross section.

An interesting case is the fermionic DM model with a light neutral pseudo-scalar mediator~\cite{Berlin:2014tja,Boehm:2014hva,Arina:2014yna,Abdullah:2014lla,Martin:2014sxa,Berlin:2014pya,Berlin:2015wwa}, in which  the Lagrangian can be written as
\bea
{\mathcal L}_s \supset  g_\chi \bar{\chi} i \gamma_5 \chi \phi + g_f \bar f i\gamma_5 f \phi \ \ \ ,
\eea
where $\chi$ is DM, $f$ represents SM fermions and $\phi$ is a pseudo-scalar. 
The direct detection cross section in this case is purely spin-dependent, and thus, the direct detection rate is significantly reduced. 
Thus, pseudo-scalar mediated DM models have received much attention in the context of explaining the GCE, with the annihilation channels $\bar \chi \chi \to f\bar f$~\cite{Berlin:2014tja,Boehm:2014hva,Arina:2014yna}.
However, there are some tensions with this model. 
First, the requirement of the correct relic density $\Omega_\chi$ prefers a moderate value of $g_f$,
which is constrained by collider searches~\cite{Kozaczuk:2015bea,Dolan:2014ska}. 
Second, in many  ultra violet (UV) complete scenarios, $\phi$ is degenerate with a  CP-even scalar boson, which is highly constrained by the LHC   heavy  Higgs searches.
It is not easy to obtain a light pseudo-scalar and heavy real-scalar with large mass splitting.
More importantly, the latest results on the dwarf spheroidal galaxies~\cite{Ackermann:2015zua} put strong bounds on the two-body fermion final states, including $\bar bb$ and $\bar\tau\tau$ final states, creating tension between the GCE signal parameter region and the allowed dwarf spheroidal region.

The above issues might be addressed when $\phi$ is lighter than $\chi$. 
In this case  $\chi$ can annihilate into a pair of $\phi$, which subsequently decay into SM particles: $\bar \chi \chi \to \phi \phi \to f\bar ff\bar f$~\cite{Abdullah:2014lla,Martin:2014sxa,Berlin:2014pya,Berlin:2015wwa}. 
In this way,  $\Omega_\chi$ does not depend on interactions between $\phi$ and $f$, which can be quite weak. 
Furthermore, the DM annihilation products are four fermion final states, which is still compatible with current constraints from dwarf spheroidal galaxies. 
This is the so-called {\it hidden sector dark matter} scenario~\cite{Feng:2008mu,Cheung:2010gj}. 
In the hidden sector scenario, there are large couplings among the hidden sector particles but small couplings of the hidden particles to the SM particles.
This scenario can evade the tension between tight constraints from direct detection and LHC searches and large GCE signature.

\begin{figure}[!htb]
\begin{center}
\includegraphics[width=0.7\textwidth]{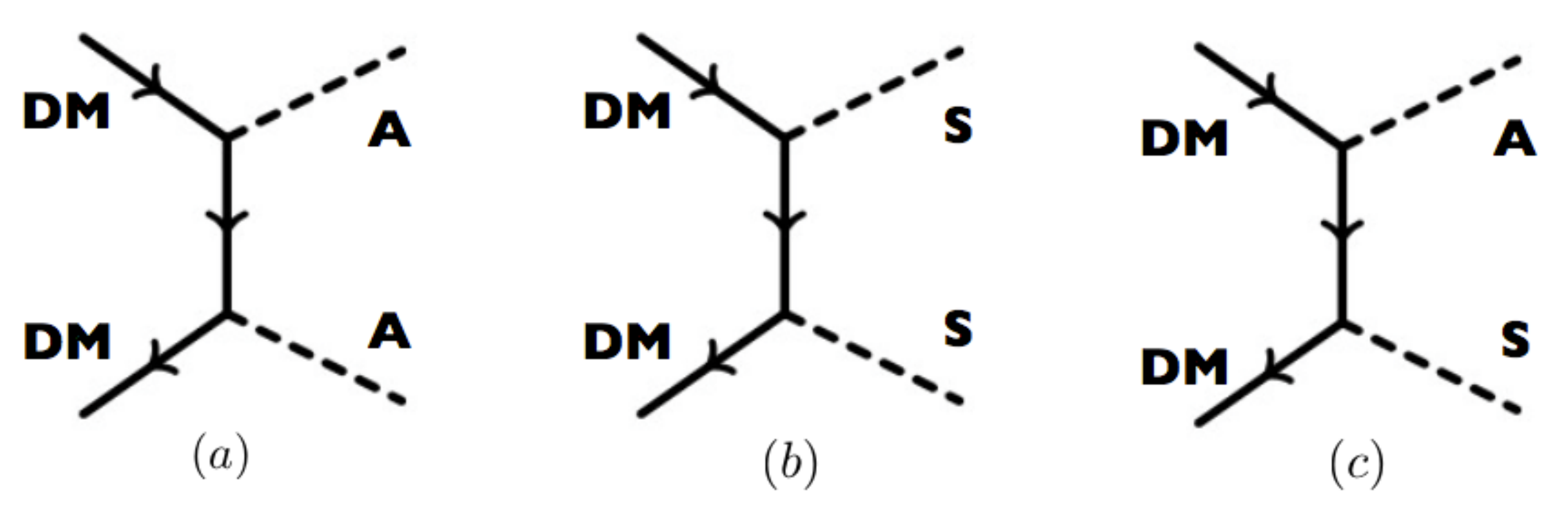}
\caption{\small The Feynman diagrams on DM annihilation to two light pseud-scalar $AA$ (a), two light real scalar $SS$ (b), and one light real scalar $S$ and another light pseudo-scalar $A$ (c). 
} 
\label{fig:spwave}
\end{center}
\end{figure}

However, there are caveats in this particular hidden sector scenario: the $\sigma(\bar \chi \chi \to \phi \phi)$ is typically $p$-wave.
This gives rise to negligible indirect detection signature.
Let us understand this from the parity transformation property of the initial and final states, and angular momentum conservation.
We know that if the annihilation amplitude has zero orbital angular momentum, the annihilation cross section should be $s$-wave annihilation. 
Under a parity (P) transformation, the fermion-antifermion initial state transforms as
$(-1)^{L+1}$, where $L$ is the total orbital angular momentum. 
Depending on the final states, we have
\bit
\item In Fig.~\ref{fig:spwave} (a), two identical pseudo-scalars in the final state. Since the two boson final state is symmetric under interchange, the P transformation are simply $P = 1$. Therefore, although total angular momentum conservation gives rise to $L = 0, 1$, from the parity we determine that the total angular momentum is $L = 1$, which implies the anninhilation cross section is $p$-wave suppressed. 
\item In Fig.~\ref{fig:spwave} (b), two identical scalars in the final state. From similar argument above, we obtain $L = 1$ and thus the anninhilation cross section is $p$-wave.
\item In Fig.~\ref{fig:spwave} (c), two different scalar bosons in the final state. In this case, there is no such exchange symmetry. Thus the orbital angular momentum could be zero. The anninhilation cross section should have $s$ and $p$-waves. 
\eit
From the above arguments, we note that if the final states have odd-number of pseudo-scalar, the annihilation is $s$-wave.
Therefore, there are several ways to realize the $s$-wave annihilation cross section in the hidden sector DM.
\bit
\item One way is that three pseudo-scalars are produced in the annihilation process, and thus the cross section $\bar \chi \chi \to \phi \phi \phi$ is $s$-wave~\cite{Abdullah:2014lla}. Although phase space suppression exists, 
if the interactions between the DM and the pseudo-scalar are much larger than   these   between the DM and the SM fermions, the annihilation channel $\bar \chi \chi \to \phi \phi \phi$ is still larger than the $s$-channel $\bar \chi \chi \to f f$. 
The gamma ray signature comes from the six fermion final states. 
However, using this channel it is  challenging to obtain both the GCE signature and the correct relic density~\cite{Abdullah:2014lla}..
\item Another possible way~\cite{Berlin:2014pya} is that if there are two light pseudo-scalars, the annihilation process $\bar \chi \chi \to \phi_1 \phi_2$ is $s$-wave. 
This still needs model building efforts to split the masses of the light pseudo-scalars from the heavy real ones.
\eit

We propose a third alternative. 
Instead of a pseudo-scalar, a complex scalar singlet  $(S=(s+ia)/\sqrt{2})$ is introduced in the hidden sector.
Due to CP violation in the scalar potential, the CP-even and CP-odd field components will mix with each other and with the Standard Model Higgs boson.
Thus,  the resulting mass eigenstates, $h_{1,2,3}$, couple to both $\bar\chi \chi$ and $\bar \chi i\gamma_5^{} \chi$ bilinears. 
Assuming $m_{h_3} \ll m_{h_2} (m_{h_1})$,  which can be realized via the CP violating terms in the Higgs potential, and $m_{h_3} < m_\chi$,
the process $\bar \chi \chi \to  h_3 h_3$  is thus kinematically allowed and the cross section can be $s$-wave. 
The reason is that the amplitude $\bar \chi \chi \to  h_3 h_3$ contains the parity odd bilinear $\bar{\chi}  i \gamma_5 \chi$.  
We show that this scenario can readily accommodate the GCE with thermal relic cross section, and still satisfy the other constraints. 
Here are the main results:
\bit
\item The annihilation rate $\bar \chi \chi \to  h_3 h_3$ depends on the CP violating phases in the scalar potential. The larger the CP violating phase, the larger of the annihilation rate in the indirect detection.
\item Direct detection depends on both the CP violation strength and the couplings of the scalars to the SM quarks. Since the hidden sector has small couplings to the SM particles,  direct detection constraints may be avoided even though there are large CP violating phases.
\item To fit the GCE spectrum, the DM annihilation cross section favors the thermal relic rate. This could be realized via sufficient CP violation. We will show that the cascade annihilation $\bar \chi \chi \to \phi \phi \to ffff$ could explain GCE while still being consistent with dwarf spheroidal constraints.
\item If $m_{h_3} < m_{h_1}/2$, the Higgs boson $h_1$ will have an exotic decay channel $h_1 \to h_3 h_3$. This gives us additional probe on the CP violating phases. Depending on the self-coupling of the complex scalar, the CP violating phases could be probed at the Run 2 LHC with high luminosity.
\eit

Our discussion of this scenario is as follows. We begin with the description of the CP violating complex scalar singlet model. In Sec. 3, we discuss the DM relic density and direction in this model. In Sec. 4, we present constraints on the model from oblique parameters and Higgs measurement. In Sec. 5, we discuss the GCE arising from the cascade annihilation.  In Sec. 6, We study signatures of the model at the LHC. We give concluding remarks in Sec. 7.


\section{The Complex CP-violating Scalar Singlet Model}
\label{sec:model}

As discussed in introduction, the hidden sector includes a Dirac fermion DM $\chi$ and 
a complex scalar singlet $S = (s+ia)/\sqrt{2}$. 
The interaction between the DM and the scalar singlet can be written as
\begin{eqnarray}
{\mathcal L}_{\rm DM} = \bar{\chi} \gamma^\mu \partial_\mu  \chi - m_0 \bar \chi_L \chi_R - y_\chi \bar \chi_L  S \chi_R + {\rm h.c.} .\label{eq:DMscalarYuk}
\end{eqnarray}
In general, the complex scalar $S$ also interacts with the SM Higgs boson.
This complex singlet scalar singlet extended SM is referred as the complex scalar singlet model (cxSM)~\cite{Barger:2008jx,Gonderinger:2012rd}. 
The tree-level scalar potential can be written as
\begin{eqnarray}
V_{\rm cxSM}&=&-\mu^2_h H^\dagger H + \lambda_h (H^\dagger H)^2-\mu_s^2 S^\dagger S +\lambda_s (S^\dagger S)^2 +\lambda_{sh} S^\dagger S H^\dagger H\nonumber \\
&&+\left[ - \mu_A^2 S^2 + \lambda_B S^2 (H^\dagger H) + \lambda_C S^4 + {\rm h.c.} \right],
\end{eqnarray}
where $H$ is the SM Higgs doublet.
Here although there is a $Z_2$ symmetry in the tree-level scalar potential, this $Z_2$ symmetry is broken by the Yukawa term in the Eq.~\ref{eq:DMscalarYuk}, and thus, there is no domain wall problem. 
The mass term $\mu_A^2$ and couplings $\lambda_B$, $\lambda_C$ can be treated as the  spurions~\cite{Haber:2012np}, which might trigger explicit or spontaneous CP violations.  We assume there is only explicit CP violation for simplicity.
To parametrize the CP-violating phases, we define the rephrasing invariants as follows: 
\begin{eqnarray}
 \delta_1 = {\rm Arg} (\lambda_B \mu_A^{2*})\; , \hspace{1cm}\delta_2 = {\rm Arg} (\lambda_C \lambda_B^{2*}) \; ,
\end{eqnarray}
 whose expressions in terms of physical parameters of the model will be given at the end of this section.
The SM Higgs doublet and $S$ are  written in component form as follows: 
$H = (G^+, v_h+\hat h+iG^0)^T/\sqrt2$,  and $S \equiv (v_s + \hat s + i \hat a)/\sqrt2$, where $v$ and $v_s$ are vacuum expectation values (VEVs) of $H$ and $S$ respectively, determined by the tadpole conditions: 
\bea
	\frac{\partial V_{\rm cxSM}}{\partial \hat \phi_i}|_{\hat \phi = ( \hat h, \hat s, \hat a)=0}   = 0.
\eea

The scalar mass matrix  in the basis $( \hat h, \hat s, \hat a)$ is
\begin{eqnarray}
{\mathcal M}^2 =\left( \begin{matrix}  2\lambda_h v^2 & (2 {\rm Re } (\lambda_B ) +  \lambda_{sh}) vv_s & -2 {\rm Im }(\lambda_B ) vv_s \\ 
\star & 2 (2 {\rm Re } (\lambda_C) + \lambda_s) v_s^2 & - 4{\rm Im } (\lambda_C) v_s^2 \\
\star & \star & 4 {\rm Re} (\mu_A^2 ) - 8 {\rm Re} (\lambda_C ) v_s^2 - 2{\rm Re} (\lambda_B) v^2   \end{matrix} \right).
\end{eqnarray}
Notice that the CP-violating couplings induce the mixings between $\hat a $ and $(\hat h,~\hat s)$.
The mass matrix can be diagonalized by the $3\times 3 $ unitary transformation $U^T {\cal M}^2 U = {\rm diag}(m_{h_1}^2, m_{h_2}^2 ,m_{h_3}^2)$, where $U$ takes the standard parametrization form and can be written as  
\begin{eqnarray}
U=\left(\begin{matrix} c_{12} c_{13} & s_{12} c_{13} & s_{13} \\ -s_{12} c_{23} -c_{12} s_{23} s_{13} & c_{12} c_{23} -s_{12} s_{23} s_{13} & s_{23} c_{13} \\ s_{12} s_{23} -c_{12} c_{23} s_{13}& -c_{12}s_{23} -s_{12} c_{23} s_{13} & c_{23} c_{13} \end{matrix}\right) \; ,
\end{eqnarray}
with $c_{ij}\equiv \cos\theta_{ij}$ and $s_{ij}\equiv\sin\theta_{ij}$. 
Then the mass eigenstates $h_i = (h_1, h_2, h_3)$ can be expressed in terms of $\hat h_i = (\hat h, \hat s, \hat a)$:
\bea
	\matThree{h_1}{h_2}{h_3} = U_{ij}^T \matThree{\hat h }{\hat s}{\hat a},
\eea
where $h_1$ is identified as the SM Higgs boson, and $h_3$ is the light mediator to the DM. 
Here $h_2$ is assumed to be very heavy to avoid possible direct detection constraints. 
The mixing angles $\theta_{13}$ and $\theta_{23}$ parametrize the CP violating phases $\delta_{1,2}$.

The DM mass is obtained as 
\bea
m_\chi = m_0 + {1\over \sqrt{2}}   y_\chi v_s, 
\eea 
and its interaction with the $S$ takes the form: 
$ y_\chi \bar \chi (\hat s + i \gamma_5 \hat a) \chi $. 
Interactions in the scalar mass eigenbasis can be parametrized as
\bea
	{\mathcal L} \supset \left[ \bar\chi (\lambda_{s i} + \lambda_{p i} i\gamma^5) \chi
	+ \bar f (g_{s i} + g_{p i} i\gamma^5) f\right] h_i,
\eea
where $f$ represents SM fermions and the coupling strengths are
\bea
	&& \lambda_{s i} = -i y_\chi U_{2i}/\sqrt2, \quad \lambda_{p i} = -i y_\chi U_{3i}/\sqrt2,\\
	&& g_{s i} = -i U_{1i} m_f/v, \quad g_{p i} = 0.
\eea
The Feynman rules for the scalar interactions are given in Table~\ref{Table:FeynRule}. 
Among these, the most relevant couplings are
\bea
	\bar\chi \chi h_3: && (-i)y_\chi(U_{23} + U_{33} i \gamma_5 )/\sqrt2; \quad 
	\bar f  f h_3:  (-i) U_{13} m_f/v; \\
	\bar\chi \chi h_1: && (-i)y_\chi(U_{21} + i U_{31}\gamma_5)/\sqrt2; \quad
	\bar f  f h_1:  (-i) U_{12} m_f/v_h.
\eea
As the hidden scalar mediator, the light mediator $h_3$ only has very small coupling to SM particles. 
This implies a small $s_{13}$   is favored. 
To have a sizable coupling $\bar\chi \chi h_3$, $s_{23}$   should be moderately large, which induces 
a large mixing between $\hat s$ and $\hat a$. 
Furthermore, to avoid constraints from the high mass Higgs searches~\cite{No:2013wsa}, the mixing angle $s_{12}$ should be small.
Finally, we need a large mass splitting between $h_2$ and $h_3$, which can be realized through the CP-violating terms in the potential. 

\begin{table}[htbp]
\begin{center}
\begin{tabular}{c|c|c|c|c}
\hline \hline  ${\rm vertex}$ &$ h_i V_\mu V_\nu$ & $h_i\bar f f$&$h_i \bar\chi \chi$& $h_i \bar \chi i \gamma_5 \chi $\\
\hline  &$ 2i U_{1i} g_{\mu\nu} m_V^2  / v_h$& $-i U_{1i} m_f/v_h$ & $-i y_\chi U_{2i}/\sqrt2$ & $ -i y_\chi U_{3i} /\sqrt2$  \\
\hline \hline
\end{tabular}
\caption{Couplings of the scalar bosons $h_i$ with SM particles.  \label{Table:FeynRule} }
\end{center}
\end{table}

Before proceeding to study constraints on the parameter space of this model, we count scalar sector physical parameters from scalar interactions, 
which are $ m_{h_1}, ~m_{h_2},~m_{h_3},~v,~v_s,~ \theta_{ij}$, $\lambda_s$ and $\lambda_{sh}$.  
The mass squared parameters $\mu_h^2 $ and $\mu_s^2$ can be determined by the tadpole conditions, while other  parameters  can be reconstructed as
\begin{eqnarray}
{\rm Re} (\lambda_B) &=&-{\lambda_{sh} \over 2}+{1 \over 2 v v_s } (m_{h_1}^2 U_{11}U_{21} +m_{h_2}^2 U_{12}U_{22} +m_{h_3}^2 U_{13}U_{23}), \\
{\rm Im} (\lambda_B) &=&-{1\over 2 v v_s} (m_{h_1}^2 U_{11} U_{31}+ m_{h_2}^2 U_{12}U_{32}+m_{h_3}^2 U_{13}U_{33}) ,\\
{\rm Im } (\lambda_C ) &=&-{1\over 4 v_s^2 } (m_{h_1}^2 U_{21} U_{31} +m_{h_2}^2 U_{22} U_{32} + m_{h_3}^2 U_{23}U_{33},\\
{\rm Re } (\lambda_C ) &=&-{\lambda_s \over 2 }+ {1\over 4 v_s^2 } (m_{h_1}^2 U_{21}^2 +m_{h_2}^2 U_{22} ^2 + m_{h_3}^2 U_{23}^2),  \\
{\rm Im}( \mu_A^2 )&=& -{1\over 2 } {\rm Im } (\lambda_B) v^2 - {\rm Im} (\lambda_C) v_s^2, \\ 
{\rm Re}( \mu_A^2 )&=& -{1\over 2 }  {\rm Re } (\lambda_B) v^2 - 2 {\rm Re} (\lambda_C) v_s^2 -{1\over 4 } (m_{h_1}^2 U_{31}^2 +m_{h_2}^2 U_{32}^2 +m_{h_3}^2 U_{33}^2),  \\ 
\lambda_{h~} &=&+{1\over 2 v^2}  (m_{h_1}^2 U_{11}^2 +m_{h_2}^2 U_{12}^2 +m_{h_3}^2 U_{13}^2) .
\end{eqnarray} 
The rephasing invariants can be expressed as
\begin{eqnarray}
\delta_1^{} &=& \arctan\left[ {{\rm Im} (\lambda_B^{}) \over {\rm Re} (\lambda_B^{} )}\right] -~\arctan \left[ {{\rm Im} (\mu_A^{}) \over {\rm Re} (\mu_A^{} )}\right] \; , \\
\delta_2^{} &= &\arctan\left[ {{\rm Im} (\lambda_C^{}) \over {\rm Re} (\lambda_C^{} )}\right] -2\arctan \left[ {{\rm Im} (\lambda_B^{}) \over {\rm Re} (\lambda_B^{} )}\right] \; .
\end{eqnarray}
%


\section{Relic Density and Direct Detection}
\label{sec:rd}

In the standard WIMP~\cite{Jungman:1995df} scenario, $\chi$ thermally freezes out, leaving a significant relic abundance.
In this model, $\chi$ and $h_3$
are assumed to be in the mass range of $10 \sim 100$ GeV. 
$h_2$ is assumed to be sufficiently heavy that its contribution to the relic density is negligible.
In this parameter region, the annihilation processes are $\chi \bar \chi \to ff$, $\chi \bar \chi \to WW/ZZ$ and $\chi \bar \chi \to h_3 h_3$. 
We will calculate the thermal relic cross sections in these channels.

\begin{figure}[!htb]
\begin{center}
\includegraphics[width=0.5\textwidth]{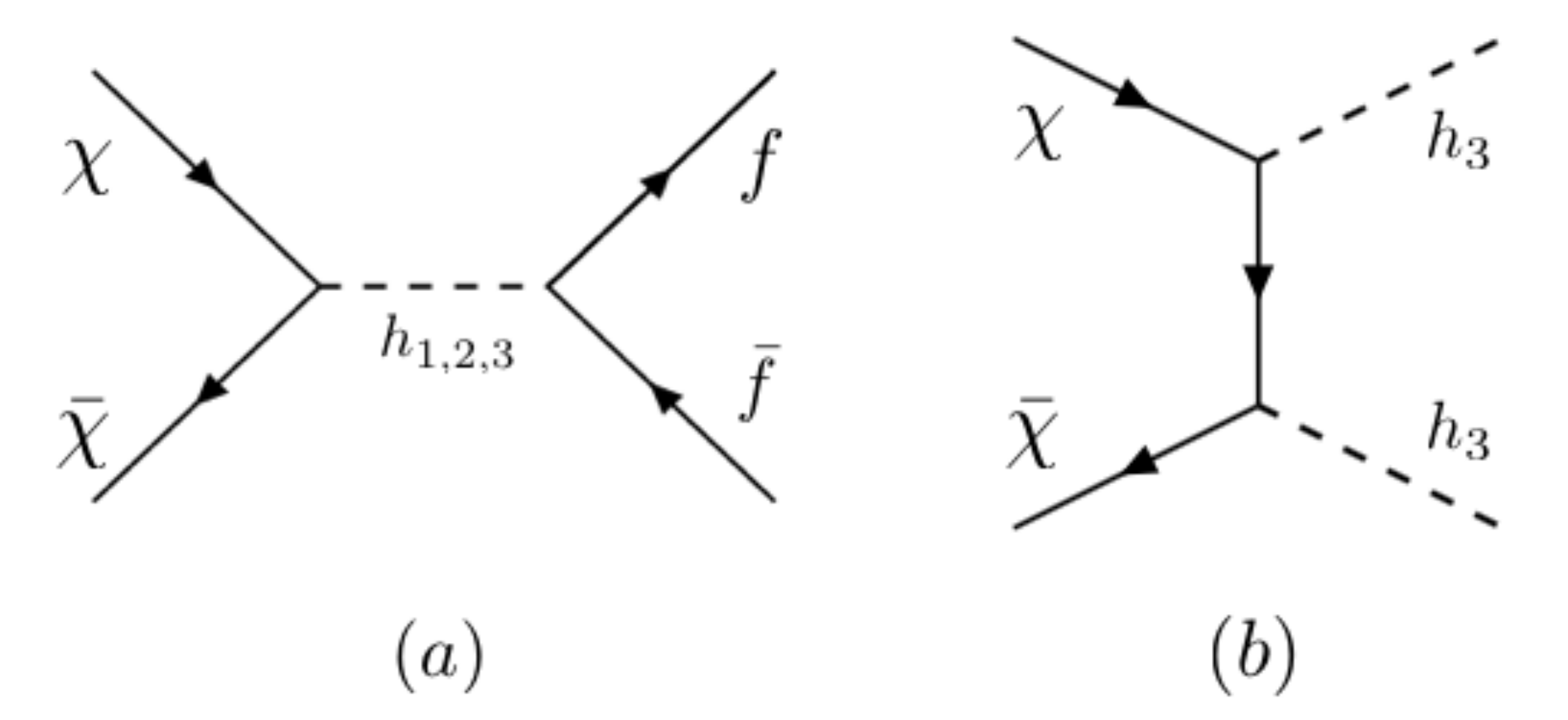}
\caption{\small Two dominant DM annihilation processes: on the left, the  $s$-channel DM annihilation process (a), and on the right, the $t$-channel DM annihilation process (b).
} 
\label{fig:relic}
\end{center}
\end{figure}

The annihilation  $\chi \bar \chi \to ff$ is through the $s$-channel $h_i$ ($i=1,2,3$) exchange, shown in Fig.~\ref{fig:relic} (a).
The $s$-wave part of this $s$-channel thermal cross section is 
\bea
	\langle \sigma v\rangle_{\bar\chi\chi \to \bar f f} &=&  N_c
	\sum_{i = 1,3} \frac{  \lambda_{pi}^2 g_{si}^2 (m_\chi^2  - m_f^2)^{3/2}}{2\pi m_\chi [(m_{h_i}^2 - 4 m_\chi^2)^2 + m_i^2 \Gamma_i^2]}, \label{schannel}
\eea
where $N_c$ is the number of colors for $f$, and $\Gamma_i$ is the total decay width of $h_i$. 
Note that this thermal cross section is proportional to the coupling strengths $\lambda_{pi}$ and $g_{si}$.
We consider that   new scalars   couple to the DM significantly but have negligible couplings to the SM particles,
which imply  small couplings $g_{si}$. Thus we expect a small thermal cross section arising from Eq. (\ref{schannel}), except in the presence of resonant  enhancements from the $s$-channel  mediators  $h_1$ or $h_3$.
When $m_\chi > m_V$, there exists the $\chi \bar \chi \to WW$ and $\chi \bar \chi \to ZZ$ channels, in which the situation is similar to that of $\chi \bar \chi \to f \bar f$.
Apart  from the resonance   enhanced  region, to obtain the correct relic density,  
the $t$-channel annihilation should be dominant over $s$-channel processes.

When $m_\chi > m_{h_3}$ and $g_{si}$ is small, the dominant channel will be the $t$-channel
process $\chi \bar \chi \to h_3 h_3$,~shown in Fig.~\ref{fig:relic} (b).
The relevant thermal cross section is 
\bea
\langle \sigma v\rangle_{\bar\chi\chi \to h_3 h_3} &=& \lambda_{s3}^2 \lambda_{p3}^2 \frac{m_\chi \sqrt{m_\chi^2 - m_{h_3}^2}}{2\pi (2 m_\chi^2 - m_{h_3}^2)^2} + (\lambda_{s3}^4 + \lambda_{p3}^4) \frac{m_\chi^3(m_\chi^2 - m_{h_3}^2)^{3/2}}{12 \pi (2 m_\chi^2 - m_{h_3}^2)^4 } \langle v^2 \rangle  \nonumber\\
&& - \lambda_{s3}^2 \lambda_{p3}^2  \frac{m_\chi^7(32 - 52 x + 20 x^2 - 3 x^3)}{48 \pi \sqrt{m_\chi^2 - m_{h_3}^2} (2 m_\chi^2 - m_{h_3}^2)^4 } \langle v^2 \rangle,
\label{eq:relicdens}	
\eea
where $x = \frac{m_{h_3}^2}{m_\chi^2}$. 
The cross section only depends on the DM couplings. 
From Eq.~\ref{eq:relicdens}, we notice that when there are both the scalar and pseudo-scalar interactions of   $h_3$ with $\chi$, the annihilation cross section could be $s$-wave. 
If the $h_3$ is purely scalar or pseudo-scalar, the annihilation cross section will be only $p$-wave.
Thus the $h_3$ needs to be a mixture of the scalar $\hat s$ and pseudo-scalar $\hat a$, which comes from the CP-violating terms in the scalar potential.

\begin{figure}[!htb]
\begin{center}
\includegraphics[width=0.45\textwidth]{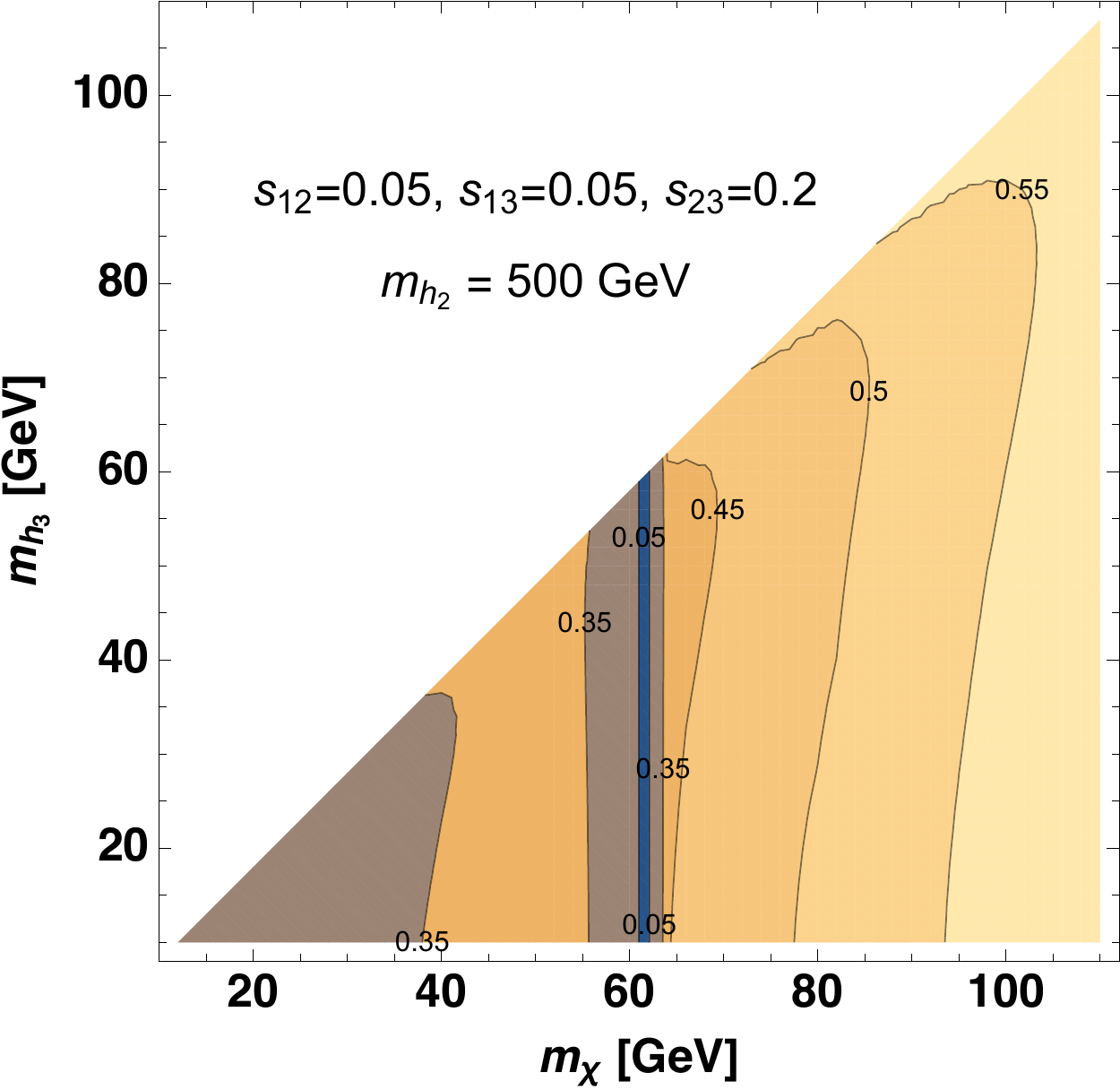}
\includegraphics[width=0.45\textwidth]{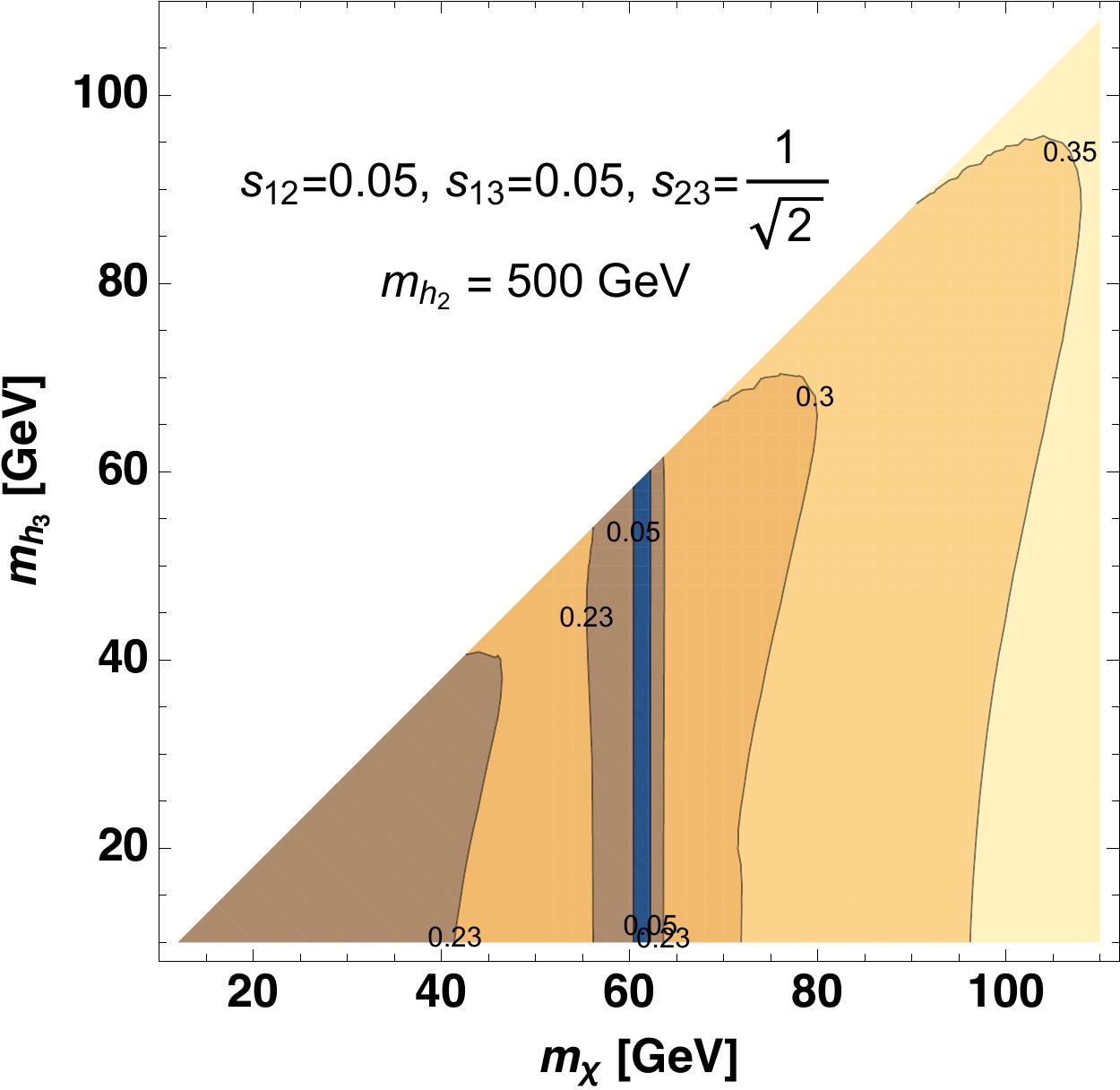}
\caption{\small Contours of the coupling strength $y_\chi$ in the $(m_\chi, m_{h_3})$ plane, which give rise to the correct relic abundance.
Mixing angles $s_{12} = s_{13} = 0.05$ and $s_{23} = 0.2$ (left panel) and $s_{23} = 1/\sqrt2$ (right panel) are assumed.  } 
\label{fig:relicmax}
\end{center}
\end{figure}

We calculate the relic density and direct detection cross section numerically using the MicrOmegas~\cite{Belanger:2006is},  which solves the Boltzmann equations numerically and utilizes CalcHEP~\cite{Belyaev:2012qa} to calculate the cross section. 
Mixing angles $s_{12}$ and $s_{13}$ need to be small to satisfy constraints from the Higgs measurements and electroweak precision test. 
On the other hand, the mixture of the $\hat s$ and $\hat a$ is parametrized by the mixing angle $s_{23}$, which we need to be large.
When $s_{23} = 1/\sqrt2$ ($\theta_{23} = \pi/4$), the mixing between $\hat s$ and $\hat a$ is maximized.
To illustrate, we choose two benchmark points: one with $s_{12} = s_{13} = 0.05$ and $s_{23} = 1/\sqrt2$, and another with $s_{12} = s_{13} = 0.05$ and $s_{23} = 0.2$. 
Given the mixing angles, we perform a parameter scan on $(m_\chi, m_{h_3}, y_\chi)$, and 
calculate the relic density for each parameter point. 
In Fig.~\ref{fig:relicmax}, we show the contours of the coupling strength $y_\chi$ in the $(m_\chi, m_{h_3})$ plane that give rise to the correct $\Omega_\chi$.
We see that away from the Higgs resonance region, $\Omega_\chi$ is dominated by the $t$-channel process. 
As the DM becomes heavier, the coupling $y_\chi$ needs to be larger to give rise to the correct $\Omega_\chi$.
This can be seen from the $1/m_\chi^2$-dependence in the thermal cross section formulae. 
Near the Higgs resonance, $m_\chi$ is almost $m_{h_1}^{}/2$. 
The $s$-channel Higgs exchange process dominates via resonant enhancement. 
Thus a small coupling  $y_\chi$ is enough to obtain the correct relic abundance.

\begin{figure}[!htb]
\begin{center}
\includegraphics[width=0.45\textwidth]{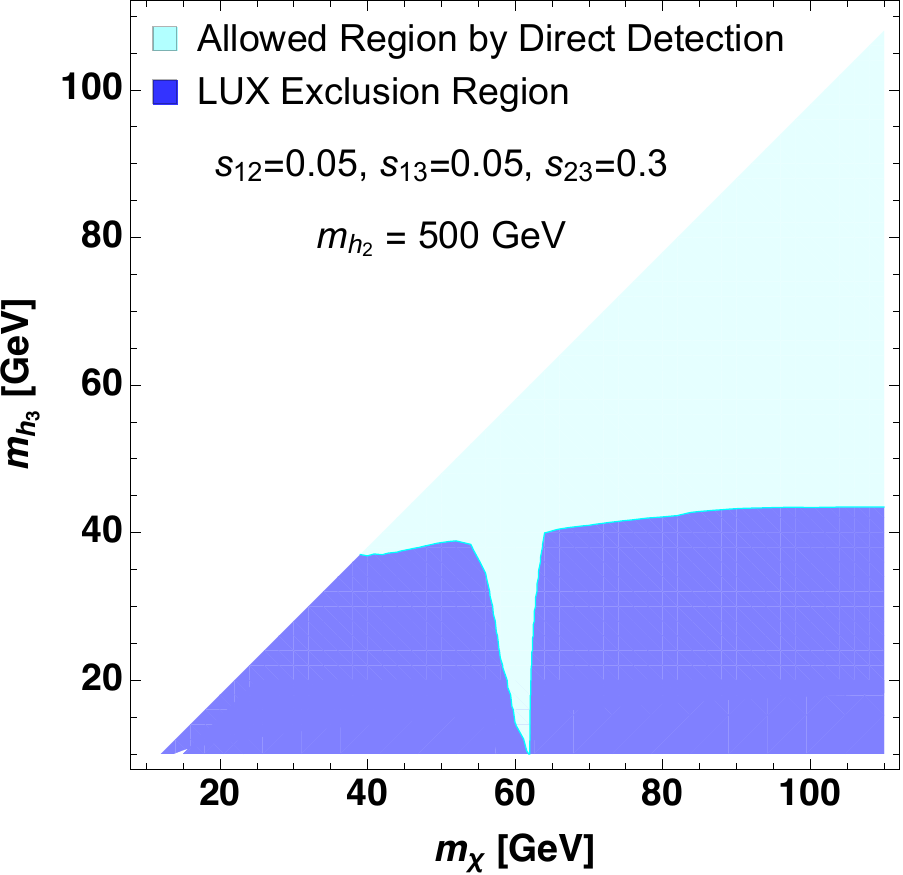}
\includegraphics[width=0.45\textwidth]{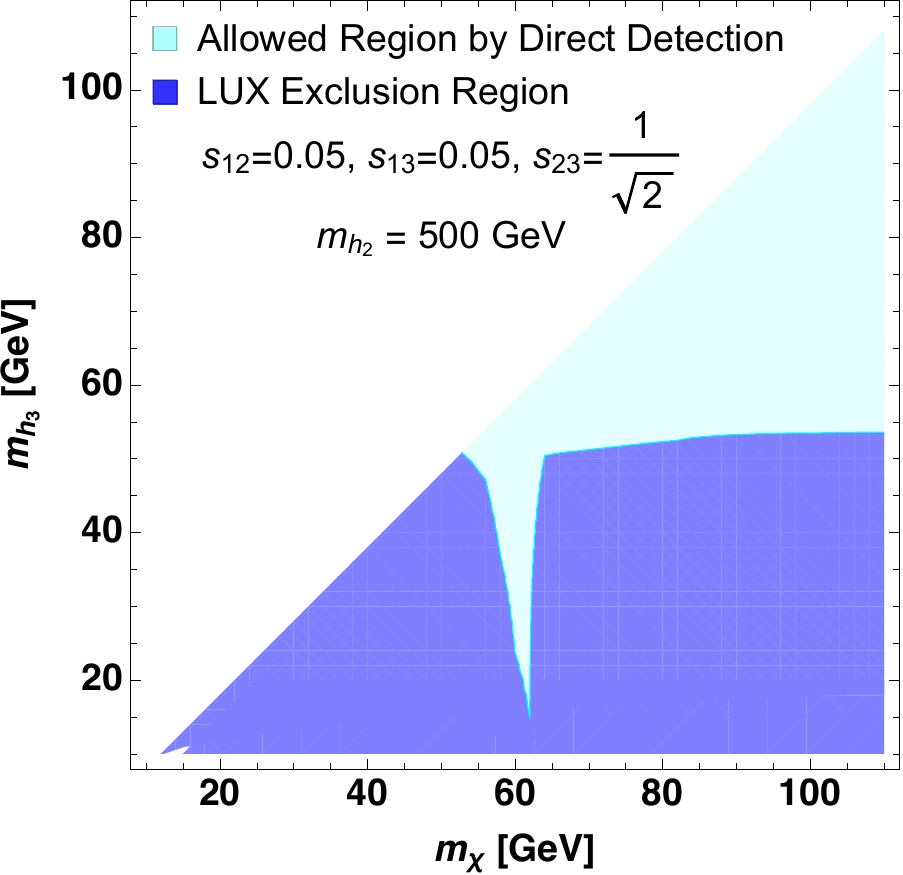}
\caption{\small Given the mixing angles $s_{12} = s_{13} = 0.05$ and $s_{23} = 0.2$ (left panel) and $s_{23} = 1/\sqrt2$ (right panel) and the coupling strength $y_\chi$ which give rise to the correct relic abundance, the blue region shows the parameter space excluded by the LUX data in the $(m_\chi, m_{h_3})$ plane. } 
\label{fig:directdmax}
\end{center}
\end{figure}

\begin{figure}[!htb]
\begin{center}
\includegraphics[width=0.2\textwidth]{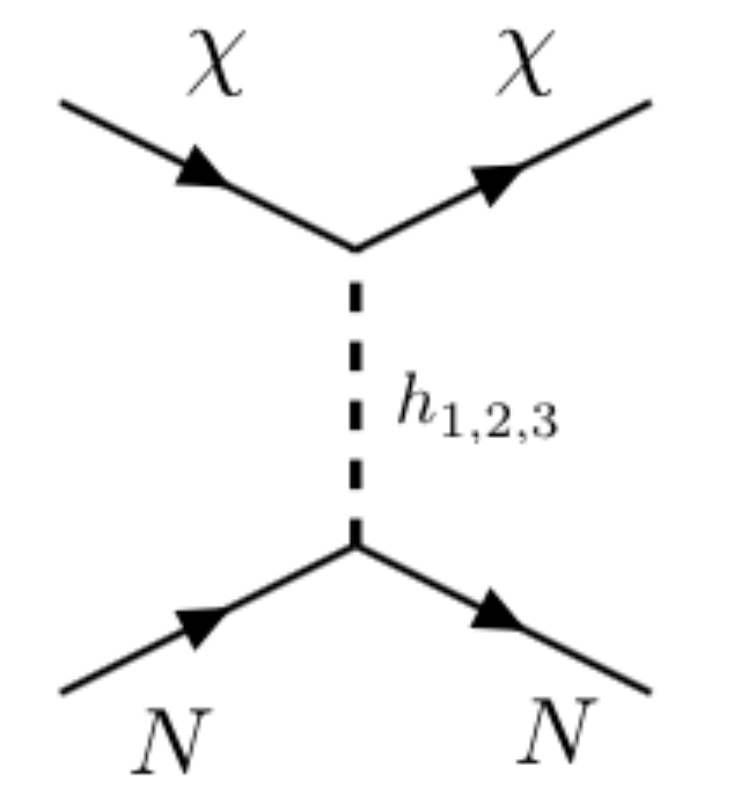}
\caption{\small DM-nucleon interaction that would generate a spin-independent direct detection signal. 
} 
\label{fig:direct}
\end{center}
\end{figure}

Given the $\Omega_\chi$-consistent parameter space, we  consider the constraints from the direct detection experiments.
Knowing that couplings of $h_i$ to the SM quarks are purely scalar type, 
the DM$-$nucleus scattering only contributes to the spin-independent (SI) scattering cross section. 
The tightest bounds on the SI cross section come from the LUX data~\cite{Akerib:2013tjd}.
In this model, the SI cross section is written as
\bea
	\sigma^{\rm SI} = \sum_{i=1,3}\frac{\mu_{\chi N}^2 U_{1i}^2 m_n^2}{\pi m_{h_i}^4 v_h^2} 
	\left(\lambda_{si}^2 +  \frac{\mu_{\chi N}^2 v^2 }{2 m_\chi^2}\lambda_{pi}^2 \right)
	\left[Z f_p + (A-Z) f_n\right]^2,
	\label{eq:sixsec}
\eea
where $\mu_{\chi N}$ is the reduced mass, $f_{p,n}$ are the form factors of the proton and neutron,  $v\sim 10^{-3}$ is the velocity of the DM. 
From Eq.~\ref{eq:sixsec}, we note that although both the scalar and the pseudo-scalar interaction of the $h_3$ to the DM contribute to the SI cross section, the scalar
interaction is dominant and the pseudo-scalar interaction exhibits velocity suppression.  
Fixing the  coupling strength $y_\chi$ by $\Omega_\chi$, we calculate the SI cross sections for different $(m_\chi, m_{h_3})$. 
Fig.~\ref{fig:directdmax} shows the exclusion limit on the $(m_\chi, m_{h_3}, y_\chi)$ parameter space, given the central value of the observed $\Omega_\chi$. 
From the Fig.~\ref{fig:directdmax}, we see that away from  the Higgs resonance region the mediator $h_3$ cannot be very light. This can be seen from the $1/m_{h_3}^4$-dependence in the SI cross section.
Near the Higgs resonance, the coupling strength $y_\chi$ is quite small, and thus the $m_{h_3}$ mass could be light.


\section{Electroweak Precision and Higgs Constraints}
\label{sec:higgs}

Typically the non-observation of permanent electric dipole moments (EDMs)~\cite{Engel:2013lsa} of neutral atoms, molecules, neutron and electron place severe constraints on the strengths of CP violation. 
In our model, these constraints are negligible, because CP violation in the scalar potential can not lead to any pseudo-scalar type Yukawa interaction of the SM fermion, which plays a  key role in generating nonzero EDMs via the two-loop Barr-Zee diagram~\cite{Barr:1990vd}. 
Constraints mainly come from the LHC Higgs measurements,   electroweak precision measurements and DM direct detections. 

\begin{figure}[t!]
\centering
\includegraphics[width=0.45\textwidth]{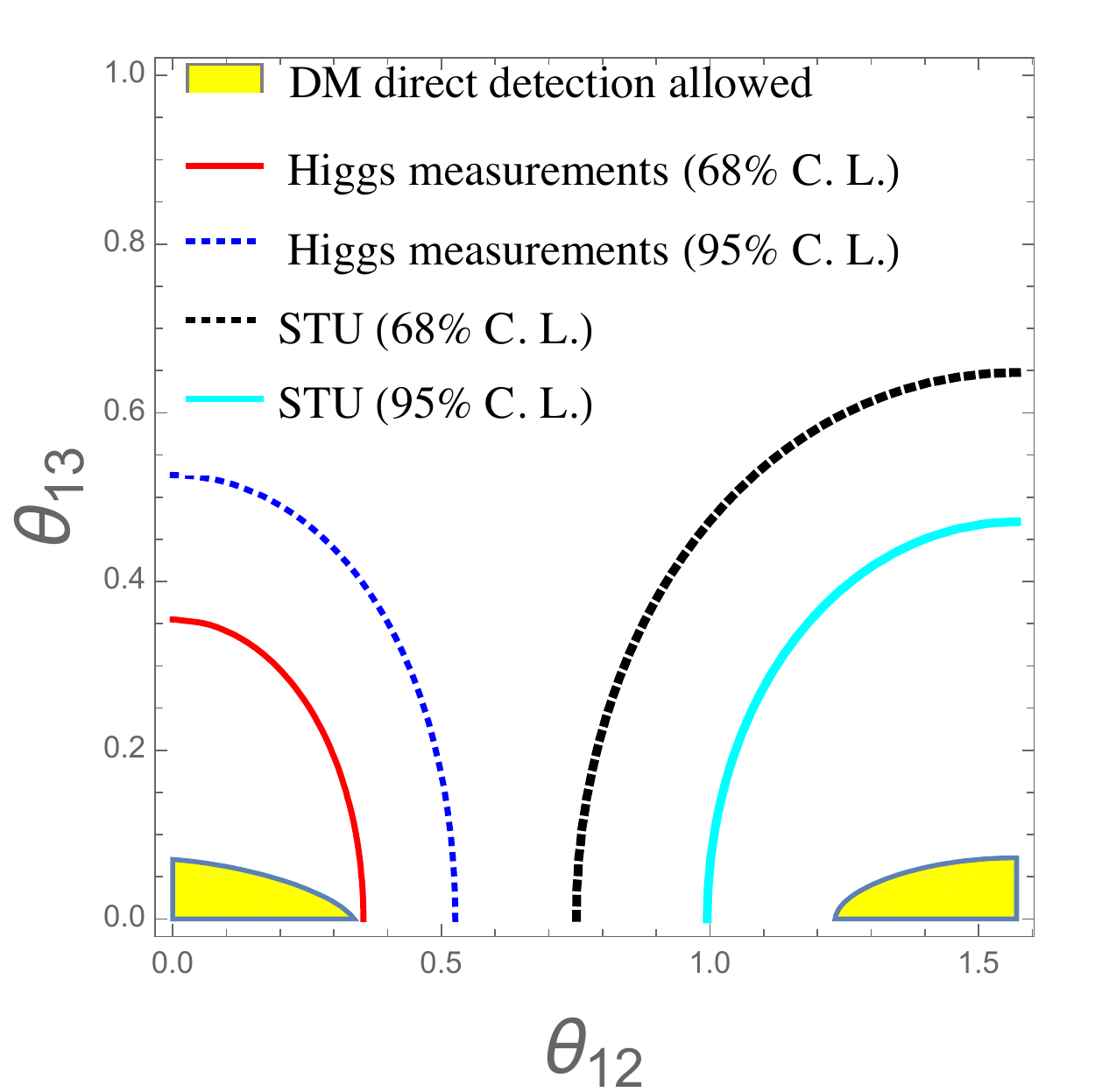}
\caption{  Constraints on the mixing angles from the Higgs measurements, oblique parameters and the DM direct detection searches. }
\label{constraints}
\end{figure}

As can be seen from the Table.~\ref{Table:FeynRule}, couplings of the SM-like Higgs to all SM particles are rescaled by the factor $c_{12} c_{13}$, the square of which equals to signal rates $\mu_{hXX}$ associated with Higgs measurements relative to SM Higgs expectations.  
In this section, we independently perform the universal Higgs fit~\cite{Giardino:2013bma} to the Higgs data from both ATLAS~~\cite{atlascomb} and CMS~\cite{Chatrchyan:2013zna,Chatrchyan:2013mxa}, where couplings of $h_1$ to paris of $t,~b,~\tau,~W,~Z,~\gamma$ equal to $r_t,~r_b,~r_\tau,~r_W,~r_Z,~r_\gamma$ in units of the SM Higgs couplings.  The $\chi^2 $  is a quadratic function of $\varepsilon_i$, where $\varepsilon_i\equiv r_i-1$, and can be written as
\begin{eqnarray}
\chi^2 =\sum_{i,j} (\varepsilon_i-\mu_i )(\sigma^2)^{-1}_{ij} (\varepsilon_j-\mu_j) \; ,
\end{eqnarray}
where $\mu_i$ are the mean value of $\varepsilon_i$, $\sigma^2_{ij} =\sigma_i \rho_{ij}\sigma_j$ with $\sigma_i$ the error of $\varepsilon_i$ and $\rho$ the correlation matrix.  
The result is shown  in  Fig.~\ref{constraints}. 
The red solid and blue dotted lines correspond to constraints at the $68\%$ and $95\%$ CL respectively.

We consider the electroweak precision constraints, utilizing bounds on the oblique observables~\cite{Peskin:1990zt,Peskin:1991sw}, which are defined in terms of contributions to the vacuum polarizations of gauge bosons.  
The dependence of $S$ and $T$ parameters  on the new scalars can be approximately expressed by the following one-loop terms~\cite{Grimus:2008nb}
 \begin{eqnarray}
 \Delta S& =&\sum_{\kappa}^{2,3} { U_{1\kappa}^2 \over 24 \pi} \left\{ \log R_{\kappa h} + \hat G (M_\kappa^2, M_Z^2 ) -\hat G(m_h^2, M_Z^2)  \right\} \\
 \Delta T &=& \sum_{\kappa}^{2,3} {3 U_{1\kappa}^2 \over 16 \pi s_W^2 M_W^2 } \left\{ M_Z^2 \left[ \log{R_{Z\kappa} \over 1- R_{Z\kappa}} -\log{R_{Zh}\over 1-R_{Zh}}\right]\right.\\&&\hspace{2.5cm}\left.-M_W^2\left[ \log{R_{W\kappa} \over 1- R_{W\kappa}} -\log{R_{Wh}\over 1-R_{Wh}}\right]\right\}
 \end{eqnarray}
where $c_W=\cos \theta_W $ with $\theta_W$ the weak mixing angle, $R_{\zeta\xi}\equiv M_{\zeta}^2/M_\xi^2$ and
\begin{eqnarray}
\hat G(M_\zeta^2,~M_\xi^2) &\equiv& -{79\over 3} +9 R_{\zeta \xi} -2 R^2_{\zeta \xi } + (12-4R_{\zeta\xi} +R_{\zeta \xi}^2 ) \hat F_{\zeta\xi} \nonumber \\
&&+ (-10+18 R_{\zeta \xi } - 6 R_{\zeta \xi}^2 + R_{\zeta \xi}^3 +9 { 1 + R_{\zeta \xi} \over 1- R_{\zeta \xi }} ) \log R_{\zeta \xi} 
\end{eqnarray} 
with
\begin{eqnarray}
\hat F_{\zeta \xi} = \begin{cases} \sqrt{R_{\zeta \xi } (R_{\zeta \xi } -4)} \log{ R_{\zeta\xi} -2 -\sqrt{R_{\zeta\xi}^2-4 R_{\zeta\xi}}\over 2} \hspace{0.5cm}\Leftarrow R_{\zeta \xi  }> 4 
\\ 0 \hspace{0.5cm}\Leftarrow R_{\zeta \xi  }=4 \\
2\sqrt{4R_{\zeta \xi } -R^2_{\zeta \xi}} \arctan \sqrt{ 4R^{-1}_{\zeta \xi } -1} \hspace{0.5cm} \Leftarrow R_{\zeta \xi }<4 \end{cases}.
\end{eqnarray}
The most recent electroweak fit  (by setting $m_{H,ref}=126~{\rm GeV}$ and $m_{t,ref}=173~{\rm GeV}$) to the oblique parameters performed by the Gfitter~\cite{Baak:2012kk} group yields
\begin{eqnarray}
S=\Delta S^0 \pm \sigma_S =0.03\pm0.10 \; \hspace{0.5cm} T =\Delta T^0 \pm \sigma_T = 0.05\pm 0.12 \; .  \label{stexp}
\end{eqnarray}
Constraints can be derived by performing $\Delta \chi^2 $ fit to the data in Eq. (\ref{stexp}), where the $\Delta \chi^2$ is given as
\begin{eqnarray}
\Delta \chi^2 =\sum_{ij}^2 (\Delta {\cal O}_i  - \Delta {\cal O}_i^0) (\sigma_{ij}^2)^{-1} (\Delta {\cal O}_j -\Delta {\cal O}_j^0) \; ,
\end{eqnarray}
in which  ${\cal O}_1 =S$, ${\cal O}_2 =T$ and $\sigma^2_{ij} =\sigma_i \rho_{ij} \sigma_j$ with $\rho_{11} =\rho_{22}=1$ and $\rho_{12} =0.891$~\cite{Baak:2012kk}. 

We show in  Fig.~\ref{constraints} oblique parameter constraints on the mixing angles  in the $\theta_{12}-\theta_{13}$ plane, by setting $m_{h_3} =60~{\rm GeV}$ and $m_{h_2} =500~{\rm GeV}$. 
The cyan solid and black dotted lines correspond to exclusion limits at the $95\%$ and $68\%$ CL, respectively. 
It is clear that the   oblique parameter constraints are much weaker than the ones from Higgs measurements. 
It is worth mentioning that the constraint of oblique parameters could be stronger by varying initial inputs of $m_{h_2,h_3}$.   

As was discussed in the last section, DM direct detection experiments also constrain the size of mixing angles.
By requiring the DM-nucleus scattering cross section to lie below the  exclusion limit put by the LUX experiment, one gets the yellow allowed region in the $\theta_{12} -\theta_{13}$ plane in Fig.~\ref{constraints}, where we have assumed $m_\chi\approx70~{\rm GeV}$, $m_{h_3}\approx 60~{\rm GeV}$, $m_{h_2} \approx 500~{\rm GeV}$ and $Y_\chi \approx 0.28$ so as as to give rise to a correct relic density.  
The direct detection cross section is approximately proportional to to $(1-2\cos 2\theta_{23})$, such that there two regions allowed in Fig.~\ref{constraints}.
The reasoning of the $\theta_{13}$ being sensitive to the DM direct detection is that we set a small  $m_{h_3}$, which is crucial for explaining the GCE.


\section{Galactic Center GeV Gamma-Ray Excess}
\label{sec:gev}

Although the direct detection and Higgs measurements put constraints on the model parameters,
there still exist large parameter regions that yield indirect detection signatures. 
Indirect detection experiments search for the products of the DM annihilation or decay.
Unlike the DM annihilation during the freeze-out, only the DM annihilation process 
with $s$-wave contribution contributes to the indirect detection signature. 
Off the Higgs resonance, the dominant $s$-wave contribution comes from the $\chi\bar\chi \to h_3 h_3$.
To see the indirect detection signature,  
the $h_3 \bar f f$  coupling  cannot be zero, and so $h_3$ could decay to the SM particles.
Thus, CP violation is needed to have indirect detection signature.
From previous sections, we learn that the $h_3$ has very small coupling to the SM fermion.
This does not affect the annihilation cross section when the $h_3$ is produced on-shell  because the branching ratio of $h_3$ does not depend on the $h_3 \bar f f$ coupling. 
Therefore, the indirect detection signatures mainly come from the cascade decay of the two on-shell $h_3$: the four-fermion final states via $\chi\bar\chi \to h_3 (\to ff) h_3 (\to ff)$. 
Among these final states, the dominant decay channel will be four-$b$ final states, because of the relatively large bottom Yukawa coupling.
This cascade annihilation and subsequent shower and hadronization produce various measurable signatures, such as gamma ray, {\it etc}.  
In the following, we will study the gamma ray spectrum of this cascade annihilation.

\begin{figure}[!htb]
\begin{center}
\includegraphics[width=0.25\textwidth]{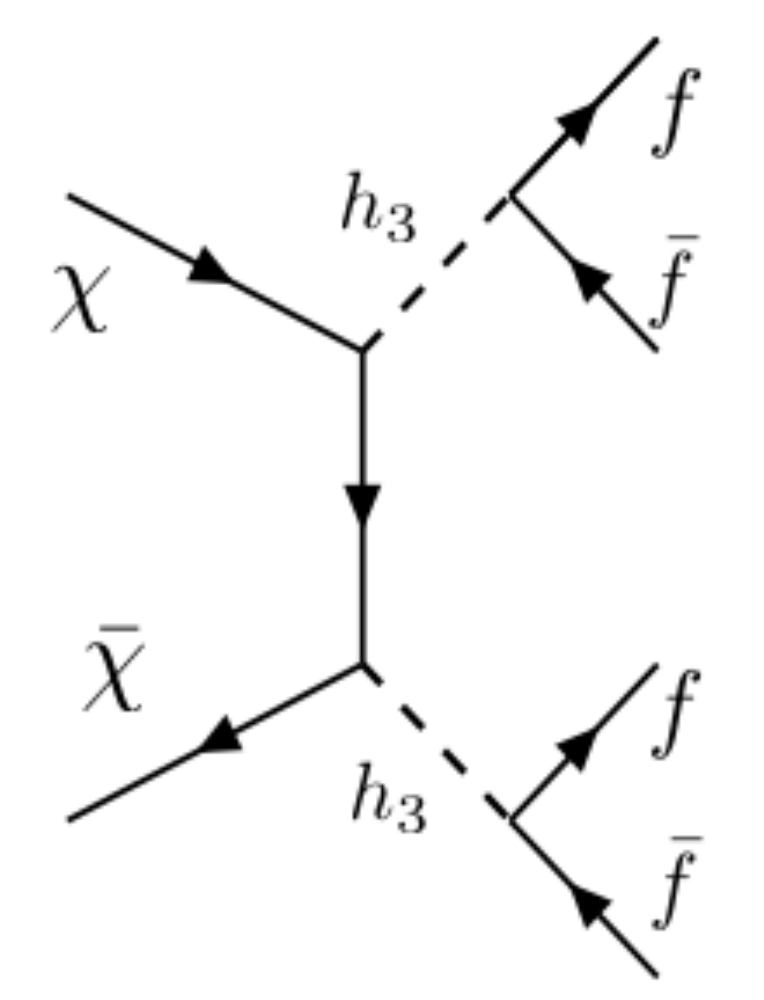}
\caption{\small  DM annihilation leading to the indirect detection signatures. 
} 
\label{fig:indirect}
\end{center}
\end{figure}

\begin{figure}[!htb]
\begin{center}
\includegraphics[width=0.45\textwidth]{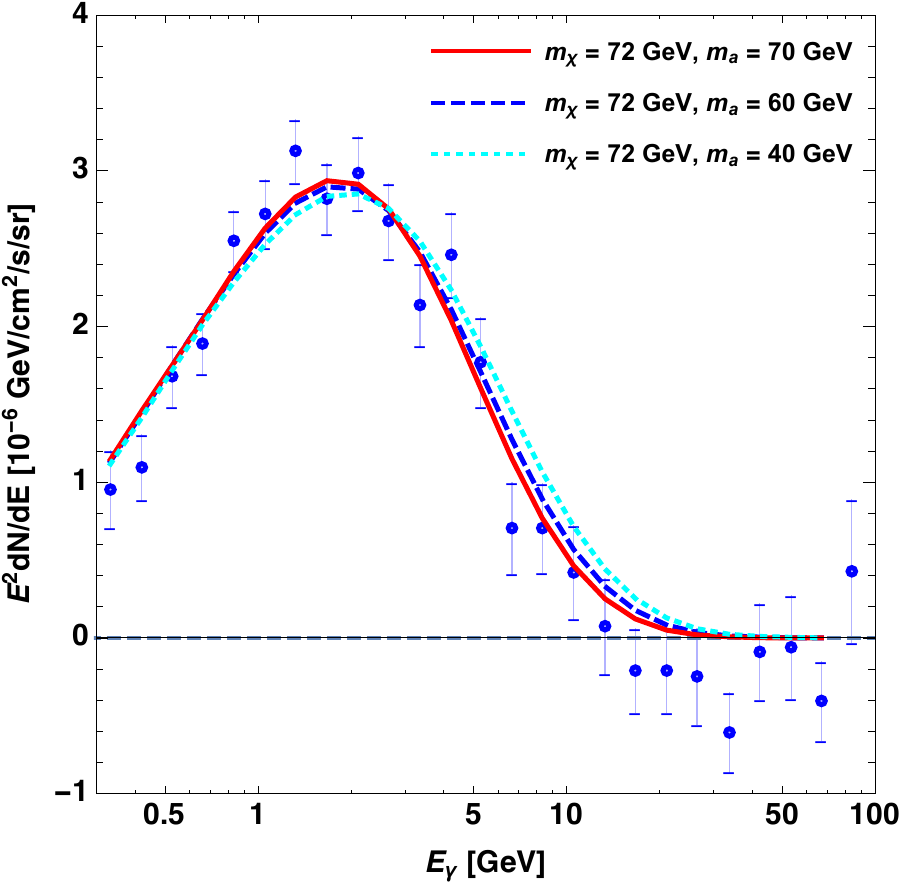}
\includegraphics[width=0.45\textwidth]{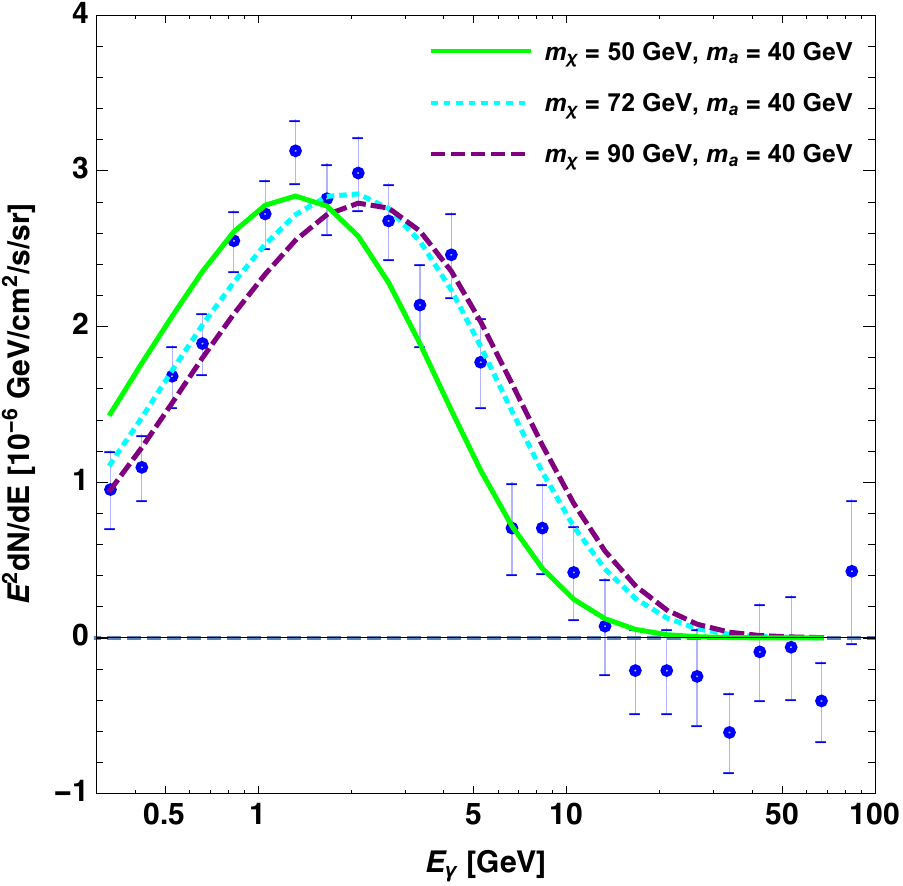}
\caption{\small The gamma-ray spectrum for different mediator masses with fixed DM mass (left panel) and 
different DM masses with fixed mediator mass (right panel). The data are taken from the spectrum of the gamma-ray excess observed in Galactic Center in Ref.~\cite{Daylan:2014rsa}. } 
\label{fig:gammaspec}
\end{center}
\end{figure}

Consider a cascade annihilation $\chi\bar\chi \to \phi (\to ff) \phi (\to ff)$, where
$\phi$ is a on-shell mediator in general. 
The gamma-ray spectrum can be obtained from the boost of the gamma-ray spectrum $\frac{dN_\gamma (\phi \to ff)}{dE_\gamma} $  in the 
$\phi$ rest frame. 
This spectrum can be easily obtained from the PYTHIA~\cite{Sjostrand:2014zea}. 
After boosting into the lab frame, the gamma-ray spectrum is written as~\cite{Berlin:2014pya}
\bea
	\frac{dN_\gamma (\chi\bar\chi \to \phi\phi)}{dE_\gamma} 
	= \frac{1}{2\beta \gamma} 
	\int^{E_\gamma \gamma (1-\beta)}_{E_\gamma \gamma (1+\beta)}
	\frac{d E'_\gamma}{E'_\gamma} 
	\frac{dN_\gamma(\phi \to ff)}{dE'_\gamma},
\eea 
where $\beta = (1 - \gamma^{-2})^{1/2}$ with the boost factor $\gamma = m_\chi/m_\phi$.
Finally, we arrive at the photon flux from the DM annihilation
\bea
	 \frac{d\Phi (b,\ell)}{dE_\gamma} 
	= \frac{\langle \sigma v \rangle_{\chi\bar\chi \to \phi\phi}}{2} 
	\frac{1}{4\pi m_\chi^2} \sum_f {\rm Br}_{\phi \to ff} \frac{dN_\gamma(\phi \to ff)}{dE_\gamma} 
	\int_{\rm LOS} d x \rho^2(r(b,\ell,x)),
\eea 
where $r(b,\ell,x) = \sqrt{x^2 + R^2 - 2 x R \cos\ell \cos b}$  is the distance from the Galactic center with galactic coordinates $(b,\ell)$, and $\rho(r)$ is the DM density profile, which is commonly taken to be the generalized NFW shape~\cite{Navarro:1995iw}. Here
the J-factor is defined as $J = \int_{\rm LOS} d x \rho^2(r(b,\ell,x))$, where $\rm LOS$ denotes the light of sight integration. 
${\rm Br}_{\phi \to ff} $ is the decay branching ratio of the mediator $\phi$ to the final state $f$, which is taken to be similar to the SM Higgs branching ratio.

\begin{figure}[!htb]
\begin{center}
\includegraphics[width=0.45\textwidth]{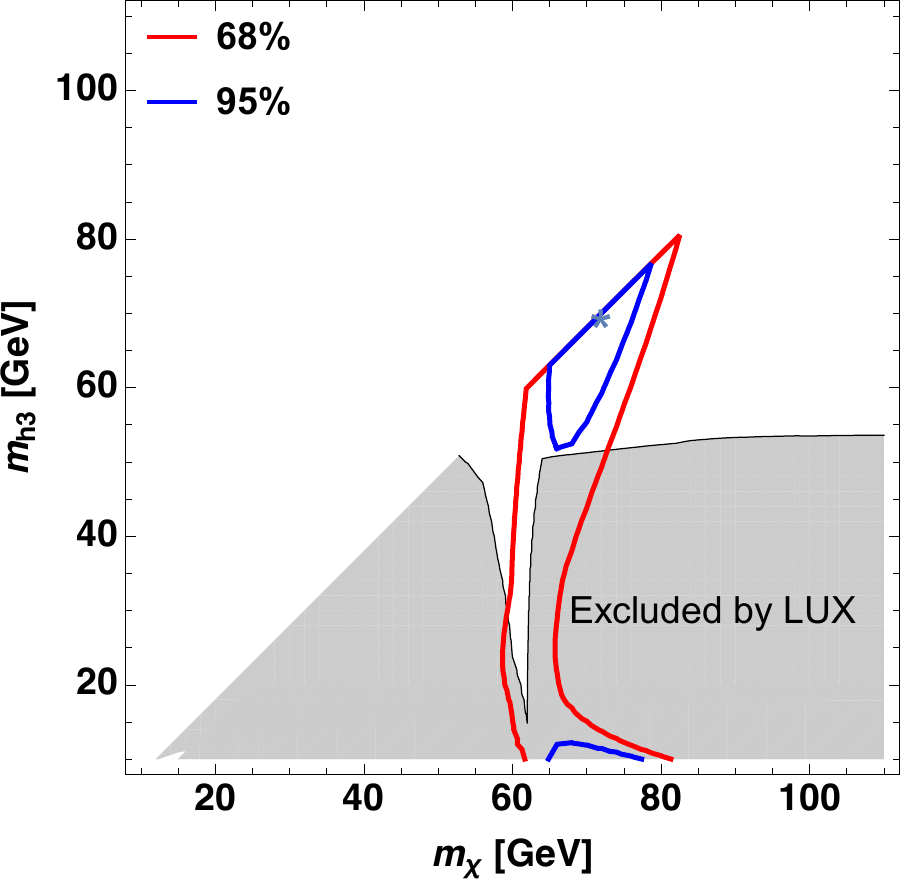}
\includegraphics[width=0.45\textwidth]{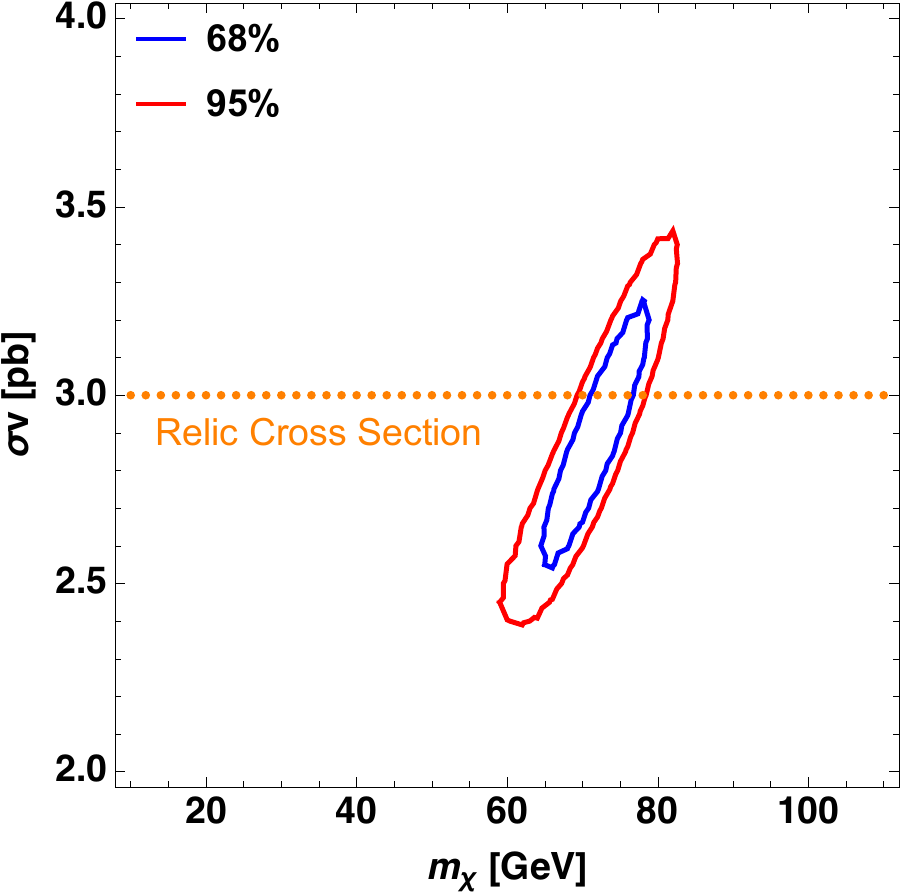}
\caption{\small On left, the favored region of the $(m_\chi, m_\phi)$ parameter space, with $\langle \sigma v \rangle$ taken to be its best value, at $68$\% and $95$\% confidence levels. The direct detection bounds (grey region) is shown, given the benchmark parameter point: mixing angles $s_{23} = 1/\sqrt2$ and 
 $s_{12} = s_{13} = 0.05$. On right panel, the favored region of the $(m_\chi, \langle \sigma v \rangle)$ parameter space, with $m_\phi$ taken to be its best value, at $68$\% and $95$\% confidence levels. 
} 
\label{fig:specfit}
\end{center}
\end{figure}

In Fig.~\ref{fig:gammaspec}, we show the gamma-ray spectrum for different $m_\chi$ and 
different $m_{h_3}$. 
The spectrum shown in Fig.~\ref{fig:gammaspec} has been normalized to corresponding to 
the J-factor $J = 9.09 \times 10^{23}$ GeV$^2/$cm$^5$ with $\gamma = 1.2$ in NFW profile.  
We find that the spectrum is very sensitive to $m_\chi$ but not to $m_{h_3}$. 
The reason is that $m_\chi$ determines the hardness of the spectrum.
To obtain the favored parameter space, we define the $\chi^2$ statistic by summing over the bins
\bea
	\chi^2  = \sum_{i} \frac{(\Phi_i^{\rm data} - \Phi_i^{\rm th}(m_\chi, m_\phi, \langle \sigma v \rangle))^2}{\sigma_i^2},
\eea
where $\Phi_i^{\rm data}$ and $\sigma_i$ are the observed flux and the error on the data given in Ref~\cite{Daylan:2014rsa}. for the bin $i$,  
and $\Phi_i^{\rm th}$ is the theoretical prediction which depends on $(m_\chi, m_\phi, \langle \sigma v \rangle)$.
Then we perform a global $\chi^2$ fit. 
In Fig.~\ref{fig:specfit} (left panel), we show the favored region of the  $(m_\chi, m_\phi)$ parameter space at $68$\% and $95$\% CLs. 
For each $(m_\chi, m_{h_3})$, the annihilation cross section is taken to be its best fit value.
We also show the direct detection bounds for one of the two benchmark choices: $s_{23} = 1/\sqrt2$,  $s_{12} = s_{13} = 0.05$.
We find that to fit with the GeV gamma-ray spectrum, $m_\chi$ is preferred to be 
around $60 \sim 85$ GeV, which is still allowed by the tight direct detection bound. 
The global fit favors the degenerate mass region for $m_{h_3}$
and $m_\chi$ with mass range $ 60 \sim 85$ GeV.
The best fit is on the parameter point $(m_\chi, m_{h_3}) = (72, 70)$ GeV.
Similarly, Fig.~\ref{fig:specfit} (right panel)  shows the favored region of the $(m_\chi, \langle \sigma v \rangle)$ plane at $68$\% and $95$\% CLs. 
For each $(m_\chi, \langle \sigma v \rangle)$, the $m_{h_3}$ is taken to be its best fit value.
We notice that the allowed annihilation cross section at  95\% CL is around $2.4 \sim 3.5$ pb,
which is the annihilation rate required by the correct $\Omega_\chi$.
Therefore, the annihilation $\chi \bar \chi \to h_3 h_3$ could simultaneously explain both the GCE and  $\Omega_\chi$. 

The gamma-ray signature at the Galactic Center is expected to appear in other galaxies, such as, dwarf galaxies. Fermi-LAT experiments investigated the Dwarf galaxies but found a null result~\cite{Ackermann:2015zua}.
This puts bounds on the gamma-ray signatures from the DM annihilation, such as
$b\bar{b}$, $\tau\tau$, and other channels. 
However, the current Fermi-LAT does not put limits on the four fermion final states with a light mediator. 
In principle, it is possible to re-analyse the Fermi-LAT data and obtain the limit on the 
four fermion final states. 
We leave this analysis for future study. 
In the following, we will comment the uncertainties on the signature and the current dwarf bounds. 
As shown in the Fig. 2 of  Ref. \cite{Ackermann:2015zua} , if the Calore, {\it et. al.} data~\cite{Calore:2014xka} are used, there is still a small parameter region which is allowed by the current dwarf bounds~\cite{Abazajian:2015raa}. 
As was pointed out in Ref. \cite{Abazajian:2015raa}, for a significantly larger integrated J-factor, which corresponds to 
an extreme high concentration/contraction Milky Way halo model, 
the signal could escape dwarf galaxy limits.

In this model, the experimental constraints implies that the hidden sector has a non-vanishing coupling to the SM sector. 
At the same time, the DM could have large coupling to the complex scalar.
It might be possible to have self-interacting DM.


\section{Hidden Scalar Searches at the LHC}
\label{sec:lhc}

Given the favored parameter space to fit the GCE signature, we would like to know
whether the LHC data are able to probe this parameter region.
Collider searches provide us a complementary way to explore the  GCE favored parameter space.
However, in this model, due to the small coupling between the SM sector and the hidden sector,
it is difficult to utilize the typical DM search channels, such as the 
mono-jet, mono-X  plus missing energy, and other pair production of the DM final states. 
Fortunately, one could still investigate the Higgs exotic decays at the LHC~\footnote{At the LEP, the $e^+e^- \to Z^* \to Z h_3$ channel is highly suppressed due to the 
small mixing angle $s_{13}$. Thus there is no constraint from the LEP data}.
If $m_{h_3}$ is lighter than half of the Higgs boson mass, 
the Higgs boson $h$ will decay to $h_1 \to h_3 h_3$.  
Similarly, if $m_\chi$ is lighter than $m_{h_1}/2$,
the Higgs invisible decay channel $h_1 \to \chi \bar\chi$ opens. Both ATLAS and CMS looked for the Higgs
invisible decay, but found null result, which puts additional constraint on the parameter space. 

\begin{figure}[!htb]
\begin{center}
\includegraphics[width=0.25\textwidth]{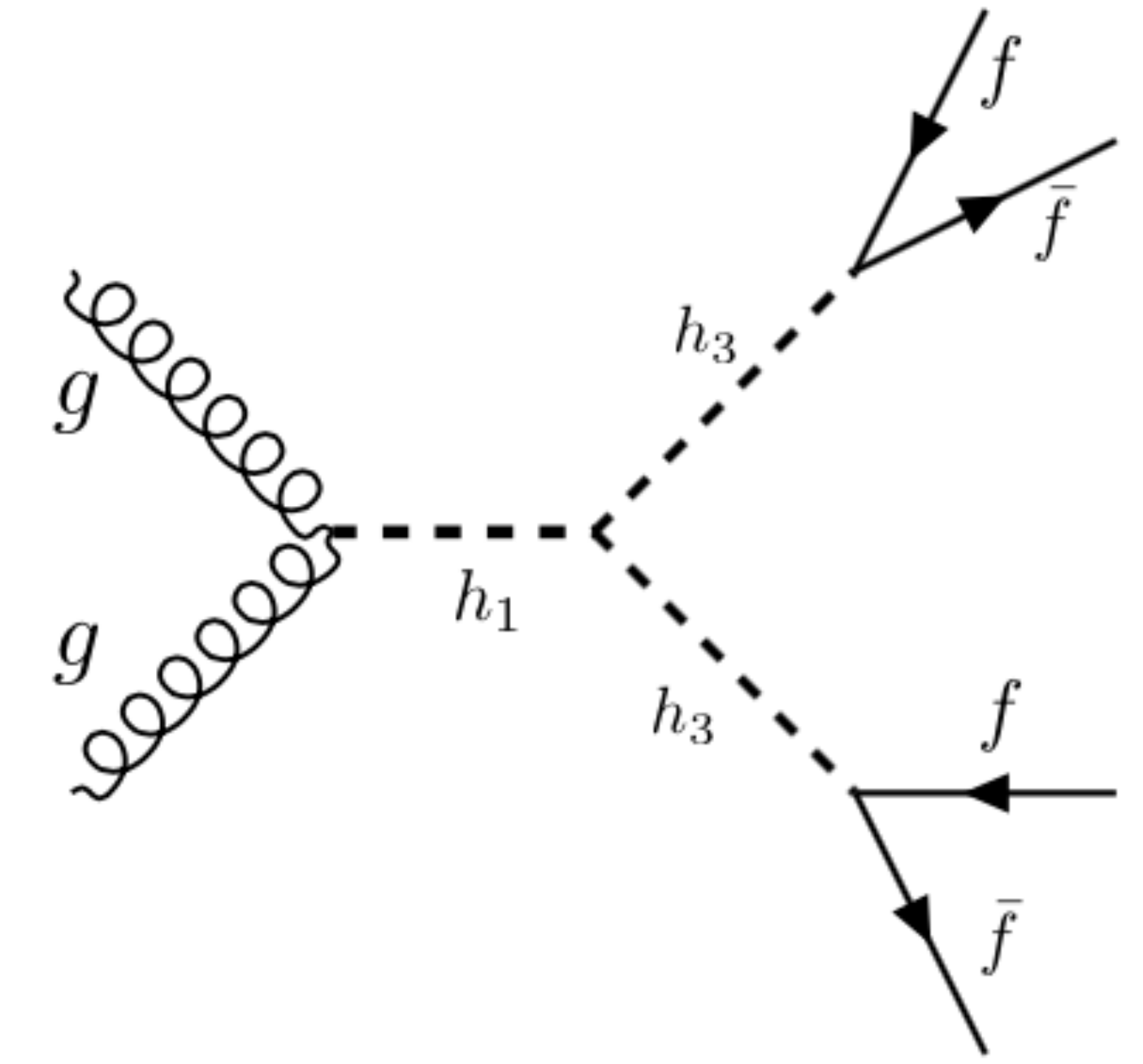}
\caption{\small  Feynman diagram for the Higgs exotic decay $h_1\to h_3 h_3 \to ffff$.
} 
\label{fig:Higgsexotic}
\end{center}
\end{figure}

Let us investigate such Higgs exotic decay rates. 
Assuming $m_h>2m_a$,  the decay rate of $h_1 \to h_3 h_3$ can be written as 
\bea
\Gamma ( h_1 \to h_3 h_3 ) \approx {\sqrt{m_h^2 -4 m_{h_3 }^2  } \left|{\cal C} \right|^2  / 32 \pi m_h^2 },   
\eea
where an extra factor $1/2$ comes from the identical particles in the final state and  the effective coupling  takes the form:
\begin{eqnarray}
{\cal C} &=& \left[ \lambda_{sh}+ 2 {\rm Re} (\lambda_B )\right ]  \left\{  v (2 U_{21} U_{23} U_{13} +U_{11} U_{23}^2) +v_s^{} (U_{21} U_{13}^2 +2 U_{11} U_{13} U_{23} ) \right \} \nonumber\\
&+ & \left [\lambda_{sh}-2 {\rm Re} (\lambda_B )\right ]   v (U_{11} U_{13}^2 +2 U_{31}^{} U_{33}^{} U_{13}^{}) \nonumber \\
&-& {\rm Im} (\lambda_B) \left\{  4v(U_{21} U_{33} U_{13} +U_{31} U_{23} U_{13} + U_{11} U_{23} U_{33})+2v_s (U_{31}^{} U_{13}^2 + 2 U_{11} U_{13} U_{33})\right\} \nonumber \\
&+ & 6 v\lambda_{h} U_{11}^{} U_{13}^2 +  \lambda_s  v_s ( 6 U_{21}^{} U_{23}^2 + 2 U_{21}^{} U_{33}^2 +4 U_{31}^{} U_{33}^{} U_{23})  \nonumber \\
&+& 12 {\rm Re} (\lambda_C)  ( U_{21}^{}  U_{23}^2  - U_{21}^{}  U_{33}^2 - 2 U_{23}^{} U_{31}^{} U_{33}^{} ) \nonumber \\
&+& 12 {\rm Im} (\lambda_C) (U_{31}^{} U_{33}^2  - U_{31 }^{} U_{23}^2 - 2 U_{23 } U_{21} U_{23} )
\label{eq:LHCscalar}
\end{eqnarray}
In the limit of small $s_{12}$ and $s_{13}$, one has
\begin{eqnarray}
\Gamma ( h_1 \to h_3 h_3 ) \approx {v^2 c_{12}^2 c_{13}^4 \sqrt{m_{h_1}^2 -4 m_{h_3}^2  }   \over 32 \pi m_{h_1}^2 } \left|\lambda_{sh}  + 2 {\rm Re} (\lambda_B)  \cos 2 \theta_{23}  -2 {\rm Im } (\lambda_B )  \sin 2\theta_{23} \right|^2  \; . 
\label{eq:hh3h3}
\end{eqnarray}
For the case $m_{h_1}> 2 m_\chi$, the Higgs to invisible decay rate can be written as
\begin{eqnarray}
\Gamma (h_1 \to \bar\chi \chi) ={ (1- c_{12}^2 c_{13}^2 )Y_\chi^2 \over 8\pi m_{H-1}^2 }  (m_{h_1}^2 -2 m_\chi^2 )^{3/2}.
\end{eqnarray}

\begin{figure}[t!]
\centering
\includegraphics[width=0.5\textwidth]{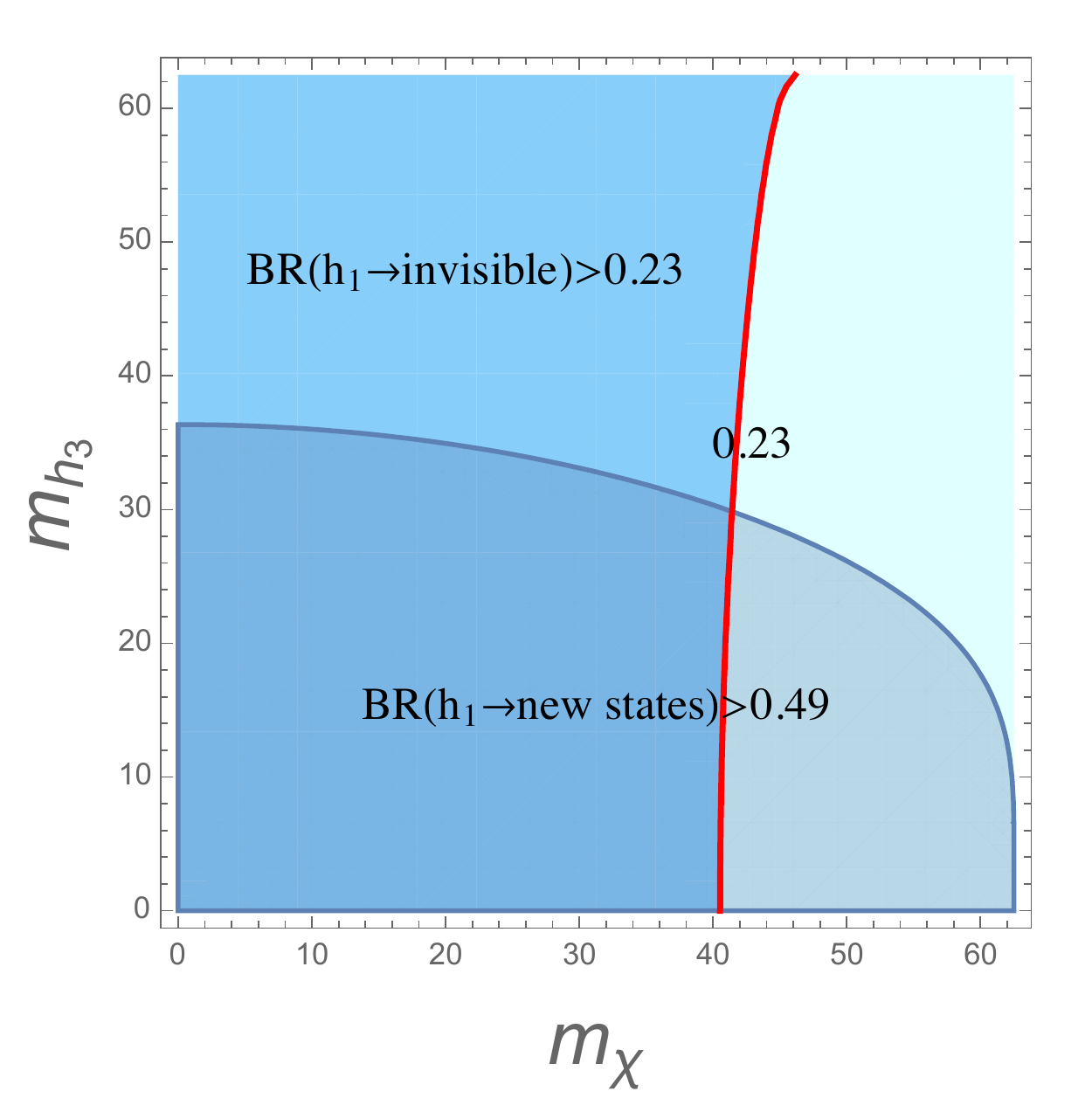}
\caption{ Contours of constant Higgs to invisible branching ratio in the $m_\psi-m_{h_3}$ plane, setting $\theta_{12}=\theta_{13}=0.05$, $\theta_{25}=0.3$, $\lambda_{sh}=0.1$, $\lambda_s=0.01$ and $y_\chi=0.4$. The red curve is the current upper bound on the Higgs to invisible decay branching ratio.  Region surrounded by the cyan curve is excluded by the upper bound on the branching ratio of Higgs to new states.  }
\label{constraintsB}
\end{figure}

The Higgs invisible decay has been studied at both the ATLAS and CMS~\cite{Aad:2015pla}. The current upper limits on the Higgs invisible decay branching ratio is ${\rm Br}(h_1 \to {\rm invisible}) < 0.23$ at the $95\%$ CL.
This limit puts constraints on the model parameters when the DM mass is lighter than the half of the Higgs boson mass.
Because $h_3$ is lighter than $\chi$, if the invisible decay channel opens,
the new exotic decay channel $h_1\to h_3 h_3$ will also exist. 
Both the invisible and new exotic channel could be classified as the undetected channel in the Higgs measurements. 
Assuming undetected channel, the Higgs coupling measurements put limit on the branching ratio of the Higgs boson decaying to invisible or undetected states ${\rm BR} (h_1\to {\rm new~states})$. 
In Ref.~\cite{Aad:2015pla}, under the assumption $\kappa_V \leq 1$ on the Higgs boson total width, ${\rm BR} (h_1\to {\rm new~states}) <0.49$ at the $95\%$ CL is obtained. 
We plot in Fig.~\ref{constraintsB} contours of the branching ratio of Higgs to invisible decay branching ratio in the $m_{\chi}-m_{h_3}$ plane by assuming  $\theta_{12}=\theta_{13}=0.05$, $\theta_{23}=0.3$, $\lambda_{sh}=0.1$, $\lambda_s=0.01$ and $y_\chi=0.4$. 
In Fig.~\ref{constraintsB}, the red curve is the current upper bound on the Higgs to invisible decay branching ratio.  Region surrounded by the cyan curve is excluded by the upper bound on the branching ratio of Higgs to new states. 
Parameter space outside two shaded regions  
satisfies the current upper limit of Higgs to invisible decays.   
From the Fig.~\ref{constraintsB}, we learn that the Higgs invisible decay could not explore the favored GCE parameter region, which favors  $m_\chi \sim 60 - 80$ GeV. 
On the other hand, the Higgs exotic decay branching ratio might be able to probe the GCE parameter region. 
In the Fig.~\ref{constraintsB}, we show that when $\lambda_s$ is small, the theoretical constraint  ${\rm BR} (h\to {\rm new~states}) <0.49$ is not sensitive to the GCE parameter region. 
In the following paragraph, we will show how the branching ratio depends on the parameter $\lambda_s$.

\begin{figure}[t!]
\centering
\includegraphics[width=0.452\textwidth]{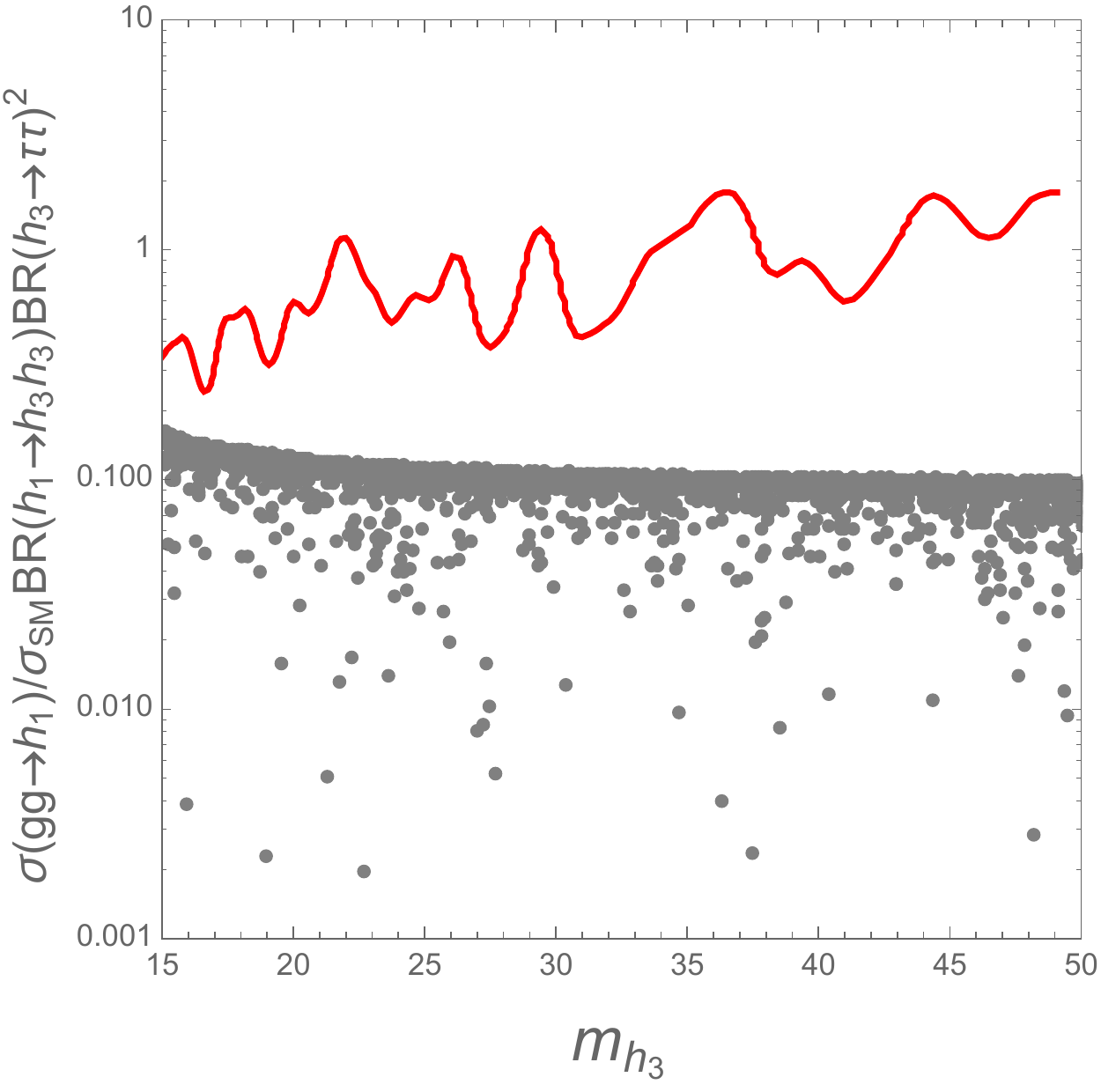}
\includegraphics[width=0.47\textwidth]{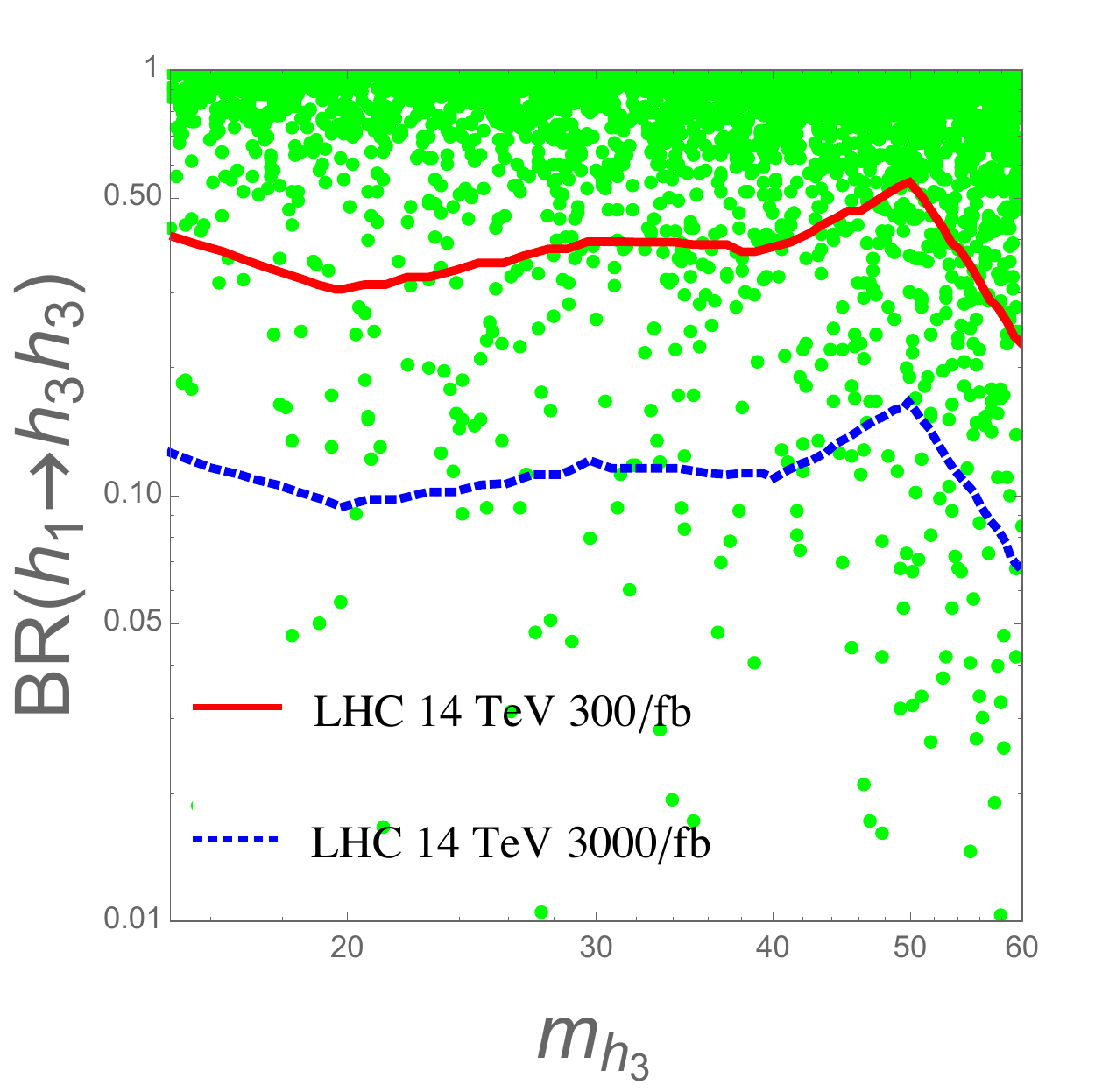}
\caption{ Scatter plots of the signal strength as the function of $m_a$. On the left, the red curve is the observed limit given by the CMS collaboration~~\cite{Aad:2015pla}.  On the right,  the red solid curve and blue dashed curves are the theoretical projections at the 14 TeV LHC~\cite{Curtin:2014pda}.}
\label{constraintsC}
\end{figure}

\begin{figure}[t!]
\centering
\includegraphics[width=0.5\textwidth]{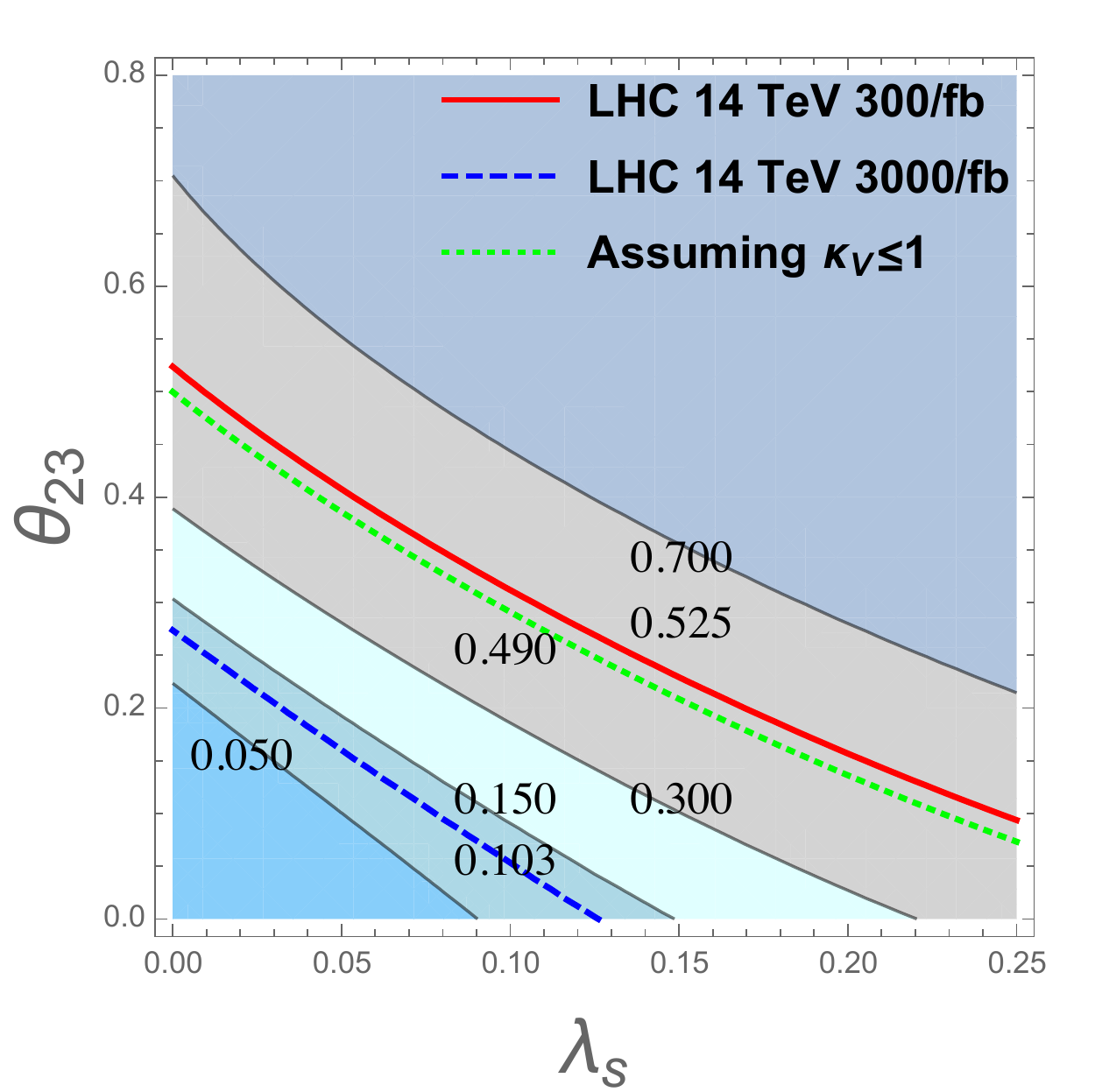}
\caption{ Given the mixing angles $\theta_{12} = \theta_{13} = 0.05$, $\lambda_{sh} = 0.1$, and the light scalar mass $m_{h_3} =  50 $ GeV, the contours of the branching ratio ${\rm Br}(h_1 \to h_3 h_3)$ in the   $\theta_{23}$ and $\lambda_s$  plane are shown. The red curve and blue dashed curve are the expected limit at the 14 TeV HL-LHC with  $300$ fb$^{-1}$ and $3000$ fb$^{-1}$ data~\cite{Curtin:2014pda}. The green short-dashed curve is the theoretical limit from the Ref.~\cite{Aad:2015pla}.}
\label{constraintsD}
\end{figure}

Except the theoretical constraints on the branching ratio of $h_1\to h_3 h_3$, 
both ATLAS and CMS also studied the exotic decay channel $h_1 \to h_3 h_3$ and set limits on the ${\rm Br}(h_1 \to h_3 h_3)$. 
The ATLAS collaboration has searched for a Higgs boson decaying into $h_3 h_3$ in the $\tau\tau \mu \mu $ channel with $\sqrt{s}=8~{\rm TeV}$ and the observed limit with the expected $\pm 1\sigma$ band was given in the Fig 6 of Ref.~\cite{Aad:2015oqa}.  
We show in Fig. \ref{constraintsC} (left) the scatter plots of the signal strength of this model as the function of the pseudo-scalar mass, which are far below the current observed limit given by ATLAS.  
With higher center of mass energy and increased luminosity at the LHC Run2, 
we expect that better sensitivity to the parameter space can be obtained.
Future LHC sensitivities have been studied in Ref.~\cite{Carena:2007jk,Chang:2008cw,Kaplan:2009qt,Cao:2013gba,Curtin:2013fra,Curtin:2014pda}. 
Ref.~\cite{Curtin:2014pda} focuses on $bb\mu\mu$ final states, and makes use of techniques of the b-tagging and the jet substructure with mass drop tagger to suppress the irreducible $bb\mu\mu$, $jj(cc)\mu\mu$, and $tt$ backgrounds. 
We utilize the projected sensitivities to ${\rm Br}(h_1 \to h_3 h_3)$ at the 14 TeV HL-LHC in Ref.~\cite{Curtin:2014pda} and recast their results to our model. 
Fig. \ref{constraintsC} (right) shows the scattering plots of the signal strength of this model as the function of the light scalar mass $m_{h_3}$. 
The red-solid and green-dashed curves show that with $300$ fb$^{-1}$ and $3000$ fb$^{-1}$ data LHC could explore most of the  parameter space in the model.

To explore the sensitivity of the GCE favored parameter region at the LHC Run2, we calculate the branching ratio ${\rm Br}(h_1 \to h_3 h_3)$ in terms of the physical parameters. 
From the decay width in Eq.~\ref{eq:hh3h3}, we note that the decay width is quite sensitive to the two parameters $\theta_{23}$ and $\lambda_s$, but not so sensitive to $\lambda_{sh}$ due to cancellation between two terms in Eq.~\ref{eq:LHCscalar}. 
In Fig.~\ref{constraintsD}, we show  the contours of  ${\rm Br}(h_1 \to h_3 h_3)$ in the   $\theta_{23}$ and $\lambda_s$  plane, given the mixing angles $\theta_{12} = \theta_{13} = 0.05$, $\lambda_{sh} = 0.1$, and the light scalar mass $m_{h_3} =  50 $ GeV.
In Fig.~\ref{constraintsD}, we note that as the branching ratio ${\rm Br}(h_1 \to h_3 h_3)$ becomes smaller, smaller values of $\theta_{23}$ and $\lambda_s$ are needed. 
Therefore, as we accumulate more data, we could probe smaller values of the parameters $\theta_{23}$ and $\lambda_s$. 
We know that to have $s$-wave DM annihilation cross section, a moderately large $\theta_{23}$ is preferred to obtain the GCE signature. 
With $300$ fb$^{-1}$ and $3000$ fb$^{-1}$ data we can reach the branching ratio to be as small as $0.52$ and $0.10$, which corresponding to $\theta_{23}$ in the region of $0.28 \sim 0.53$, which is the interesting parameter region.


\section{Conclusions}
\label{sec:conclusion}

We have investigated a hidden dark matter scenario, in which the hidden sector includes a fermion DM $\chi$ and a complex scalar $S$. 
The complex scalar mixes with the SM Higgs boson with suppressed coupling. 
This solves the possible tension between tight constraints from direct detection and LHC searches and large indirect detection signature.
To obtain large DM annihilation rate, we propose that there are the CP violations in the scalar potential. 
The CP violations could mix the real and pseudo-scalar parts of the complex scalar, and induce a large mass splitting for the mass eigenstates: a light $h_3$ and much heavier $h_2$. 
We focus on an interesting parameter region:
the light scalar $h_3$ is lighter than the DM $\chi$.  
This allows the process $\bar\chi\chi \to h_3 h_3$ as the dominant DM annihilation channel with $s$-wave cross section. 
This annihilation channel gives rise to significant indirect detection signature and could explain the existing Galactic Center gamma-ray excess.

The relevant physical parameters are the DM and light scalar masses, the mixing angles among the Higgs boson and the real and imaginary part of the complex scalar boson: $\theta_{12}$,  $\theta_{13}$, and $\theta_{23}$, the   Yukawa coupling between the DM and the complex scalar $y_\chi$. 
To obtain the needed $s$-wave enhancement of DM annihilation, we requires the mixing angle between the real and imaginary part of the complex scalar boson $\theta_{23}$ to be large.
On the other hand, the direct detection bounds implies small mixing angles $\theta_{12}$ and $\theta_{13}$, and large mass splitting between $h_2$ and $h_3$. 
We found that the EDM constraint is   negligible, but the constraints from the DM direct detection, electroweak precision and the Higgs coupling measurements are tight.
Both the electroweak precision and the Higgs coupling measurements prefer small mixing angles $\theta_{12}$ and $\theta_{13}$.
These constraints force us to consider the hidden DM scenario in this model.

To explain the Galactic Center excess, the DM annihilates into the four-fermion (mainly four-$b$) final states {\it via} the cascade decay $\chi\bar\chi \to h_3 (\to ff) h_3 (\to ff)$. 
To fit the observed gamma-ray spectrum, the $m_{\chi}$ is prefered to be in the 60 to 80 GeV region. And $m_{h_3} \simeq m_\chi $ is favored. 
Moreover, the annihilation cross section is fitted to be in the region to have correct relic density.
In short, this hidden DM model explain the gamma-ray excess.
Although the dwarf galaxies might place additional constraints, it is still possible to be compatible with the current bounds if the extreme integrated J-factor is adopted.
Because the  scalar $h_3$ cannot be too light, it is unlikely to have self-interacting DM in this hidden scalar scenario.

We also found that constraints from the Higgs invisible decay and Higgs width imply that the $\chi$ and the light scalar $h_3$ cannot be very light. This constraint is consistent with the observed gamma-ray signature.
Although the hidden sector has small coupling to the SM Higgs boson, if $h_3$ is lighter than half of the Higgs mass, the $h_1 \to h_3 h_3$ is a golden channel to investigate. 
The 8 TeV LHC results on the exotic decay $h_1 \to h_3 h_3$ cannot probe the favored parameter region. 
However, we show that the future 14 TeV studies could be sensitive to the mixing angle $\theta_{23}$, which controls the DM annihilation rate.
Thus we expect that with 300 fb$^{-1}$ and 3000 fb$^{-1}$ data the exotic decay $h_1 \to h_3 h_3$ process will be able to probe the parameter region favored by the Galactic Center gamma-ray excess.

\section*{Acknowledgements}

We thank Kevork Abazajian, Grigory Ovanesyan and Peter Winslow for helpful discussions.
JHY also thank Wei Xue and Asher Berlin for useful conversation during CETUP in South Dakota. 
This work was supported by U.S. Department of Energy contract DE-SC0011095.



\end{document}